\documentclass[aps,pre,twocolumn,superscriptaddress,longbibliography,nofootinbib]{revtex4-2}

\usepackage[T1]{fontenc}
\usepackage[utf8]{inputenc}
\usepackage{lmodern}          
\usepackage{amsmath,amssymb}  
\usepackage{graphicx}         
\usepackage{tabularray}       
\usepackage{xcolor}           
\usepackage{mathtools}

\usepackage[colorlinks=true, allcolors=blue, citecolor=blue, linkcolor=blue, urlcolor=blue]{hyperref}
\usepackage{cleveref}

\newcommand{\real}{\text{real}}
\newcommand{\argmax}{\text{argmax}}

\newcounter{commentzaehler}

\begin{document}

\title{Equivalency of spike-frequency and h-current based adaptation in a Wilson-Cowan field model}

\author{Ronja Str\"{o}msd\"{o}rfer}
\email[Corresponding author: ]{stroemsdoerfer@tu-berlin.de} 
\affiliation{Neural Information Processing Group, Fakult\"{a}t IV, Technische Universit\"{a}t Berlin, Berlin, Germany}
\affiliation{Einstein Center for Neuroscience, Berlin, Germany}

\author{Klaus Obermayer}
\affiliation{Neural Information Processing Group, Fakult\"{a}t IV, Technische Universit\"{a}t Berlin, Berlin, Germany}
\affiliation{Einstein Center for Neuroscience, Berlin, Germany}
\affiliation{Bernstein Center for Computational Neuroscience, Berlin, Germany}

\date{\today} 

\begin{abstract}
During slow-wave sleep, the brain produces traveling waves of slow oscillations (SOs; $\leq 2\ \rm{Hz}$), characterized by the propagation of alternating high- and low-activity states. They play a crucial role in memory consolidation and are frequently investigated, empirically and with the support of in-silico studies. The question of internal mechanisms that modulate traveling waves of SOs is still unanswered even though it is established that it is an adaptation mechanism that mediates them. One of the main mechanisms investigated is spike-frequency adaptation (SFA), a hyperpolarizing feedback current that is activated during periods of high-activity. An alternative mechanism which has recently been suggested is based on hyperpolarization-activated (h-)currents, which are positive feedback currents that are activated in low-activity states. Both adaptation mechanisms were shown to feature SO-like dynamics in neuronal populations, and the inclusion of a spatial domain seems to enhance observable differences in their effects. To investigate the effects of both adaptation mechanisms on the neural dynamics in detail, we examine a spatially extended, adaptive Wilson-Cowan model of interacting excitatory and inhibitory populations of neurons with local spatial coupling. The excitatory populations are equipped with either one or the other adaptation mechanism. We describe both mechanisms using the same dynamical equation and include the inverse mode of action by changing the signs of the adaptation strength and the gain of activation function. We then show that the dynamical systems including the two feedback currents are mathematically equivalent under a compensatory external input, which depends on the adaptation strength and which in turn leads to a shift in state space of the otherwise equivalent bifurcation structure. A detailed analysis of state space shows that strong enough adaptation is required to induce traveling waves of SO-like dynamics while only stationary or homogeneous activity patterns emerge if adaptation is weak. Additionally, adaptation modulates the properties of the spatio-temporal activity patterns, such as temporal frequencies, spatial frequencies, and the speed of the traveling waves, all of which increase with increasing strength. Though being dynamically equivalent, our results also explain why location-dependent variations in feedback strength cause differences in the propagation of traveling waves between both adaptation mechanisms.
\end{abstract}


\maketitle

\section{Introduction}\label{sec:introduction}
One of the most frequently investigated rhythms in the brain is the phenomenon of slow oscillations (SOs), which appear during slow-wave sleep and are characterized by the slow alternation of high- and low-activity states, also referred to as up- and down-states \cite{Sanchez-Vives2017, SanchezVivez2020}. They were first described by \cite{Steriade1993}\footnote{Note, however, that cortical down states, which correspond to so-called K-complexes, have already been described in the very first sleep EEG studies in the 1930s \cite{Loomis1935}.}, and have been shown to play a crucial role in memory consolidation since then \cite{Brodt2023, Klinzing2019, RaschBorn2013}. On a mesoscopic scale, SOs tend to travel in wave-like patterns across the cortex \cite{Massimini2004}. Due to their dominant temporal frequency being below $\leq 1\ \rm{Hz}$, SOs are most likely driven by an adaptation process which acts on a slower time scale than the standard neuronal response. Nonetheless, the exact nature and dynamics of the adaptation mechanism are still in question.

One candidate process is spike-frequency adaptation (SFA), a hyperpolarizing feedback current (i.e.,\ a negative feedback mechanism) induced by ongoing high-activity in neuronal populations which is often mediated by slow potassium (K$^+$) currents (see \cite{Faber2003, Ha2017}). Since the neurotransmitter acetylcholine (ACh) decreases K$^+$ conductance \cite{McCormick1992} and since ACh concentration is lower during natural sleep compared to wakefulness \cite{Jones2003}, SFA is elevated during sleep. Comparing different species, two types of slow waves were separated in \cite{Nghiem2020} depending on whether they occur in natural slow-wave sleep or whether they occur under anesthesia. \cite{Jercog2017} reported that the durations of up-states following down-states are positively correlated during anesthesia in rats, while \cite{Nghiem2020} observed a negative correlation, with long down-states being followed by short up-states in natural sleep in several species. In \cite{Nghiem2020}, reduced cholinergic modulation (i.e.,\  lowered ACh concentration) increased the incidence of short up-states subsequent to long down-states, thereby recapitulating both the slow-wave patterns of natural sleep and the neuromodulatory effects of varying ACh concentrations in the sleeping brain.
\vspace{-2mm}

Hyperpolarization-activated h-currents (i.e.,\ a positive feedback mechanism) have also been shown to slowly modulate neuronal excitability (see \cite{Combe2021Review}) and may also cause adaptation-driven traveling waves similar to cortical waves of SO-like dynamics. In \cite{Shu2023}, using ferret slice experiments, it was proposed that h-current modulation regulates the cortical capacity for recurrent persistent activity, which encompasses slow-wave dynamics. Until most recently, h-currents were predominantly investigated in thalamocortical systems \cite{DestexheBabloyantz1993, Destexhe1993, HillTononi2005} which aligns with their presence in thalamic neurons \cite{Abbas2006}. However, \cite{Mehrotra2024}, and \cite{Porta2024} focus on the impact of h-currents on cortical slow waves, i.e.,\ on the propagation of alternating up- and down-states over space.

Spatially extended neural field models are a convenient computational approach for investigating the propagation features of traveling waves. They inherit a description of space through spatial convolution, which effectively describes distance-dependent connectivity profiles in continuous space \cite{Bressloff2012, Ermentrout1998, Ermentrout2010mathematical_book}. Depending on their set-up, neural field models can produce periodic traveling waves \cite{Budzinskiy2020}. For the one-population model of integrate-and-fire neurons with synaptic inputs on a continuous one-dimensional space of \cite{Erazo-Toscano2023}, analytical predictions align with numerical simulations showing that the speed of traveling waves fluctuates over space when there is a subset of tightly coupled neurons. In \cite{Omelchenko2024}, traveling waves emerge in a system of quadratic integrate-and-fire neurons with nonlocal synaptic coupling as the synaptic coupling strength is increased. Generally speaking, the emergence of spatio-temporal patterns relies on the underlying dynamical state of the model system. Suitable parameterizations can be identified by a Turing instability analysis performed in the Fourier domain. This analysis is most often performed in systems comprised of either a single neural population (e.g.,\ \cite{Coombes2005, Quabbaj2009}) or two coupled populations of excitatory and inhibitory neurons (e.g.,\ \cite{Harris2018, Wyller2007}).

One frequently used model is the Wilson-Cowan model which was first introduced in \cite{WilsonCowan1972} in its mass model version, i.e.,\ without a spatial domain. The mass model version displays a rich variety of emergent dynamics and has previously been used to investigate SFA-driven SOs, either with \cite{Dimulescu2025, Levenstein2019} or without inhibition \cite{Torao2021}. In its original form, comprising of one excitatory and one inhibitory population of neurons without SFA \cite{Bressloff2010}, the dynamical landscape of this model remains fairly similar when both SFA and inhibition are included, when only SFA is present, or when both are absent. A regime of oscillations (e.g.,\ emerging due to a Hopf instability) is at the corner end of a regime of bistability, where a stable up- and down-state coexist. Modified versions of the model have been constructed that include finer biophysical detail, e.g., synaptic plasticity; while maintaining the systems dimension at two differential equations as in \cite{Trabocchi2025}, and the model has even found applications in the machine learning field \cite{Marino2025}.

To include a spatial domain, Wilson and Cowan extended the mass model to a field model which is comprised of two coupled populations of excitatory and inhibitory neurons \cite{Wilson1973}. Pinto and Ermentrout \cite{PintoErmentrout2001} investigated the system without an inhibitory population, but with a linear, negative feedback that limits the networks excitability. They establish the existence of traveling pulses in such a network and connect the speed of the waves, which are induced by the slow negative feedback, to the model parameters using the Heaviside function as transfer function. 
For a homogeneous, excitatory neural field model, \cite{Amari1977} proved that lateral inhibition is required to stabilize bump solutions (i.e.,\ stabilize the activity to not decay or vanish over time), and while \cite{Folias2004} confirms these findings, they additionally show that local inhomogeneous input stabilizes stationary-pulse solutions in an otherwise homogeneous network. In \cite{Kilpatrick2010}, the authors investigate the same field model without inhibition in which nonlinear negative feedback to an excitatory population of neurons is produced by synaptic depression and SFA simultaneously. They identify traveling fronts and pulses, as well as stationary or homogeneous oscillations. They show that synaptic depression is not sufficient to stabilize bump solutions and confirm that either lateral inhibition (as shown in \cite{Amari1977}) or external input (as shown in \cite{Folias2004}) is required. Conversely, bumps do not exist when SFA is present. 
Their nonlinear formulation of SFA feedback is a physiologically based negative feedback. In \cite{Curtu2004}, the authors equip the excitatory population of the two-population model with SFA and investigate the conditions for Turing and Hopf bifurcations to emerge, enabling stationary or spatio-temporal activity. While weaker and faster adaptation generates stationary patterns, slower and stronger adaptation causes the stationary patterns to travel.

Now, we are left with the question of what the differences are between the slow negative feedback mechanism based on SFA, see \cite{Augustin2013, Cakan2020, Cakan2022, Ladenbauer2014, Levenstein2019, Mattia2012}, and the recently proposed positive feedback mechanism based on h-currents, see \cite{Mehrotra2024, Porta2024} which have already been shown to induce SO-like dynamics. Simultaneous investigations of h-currents and SFA were already conducted. \cite{Roth2020} used fluorescent dye in the dentate gyrus of rats and unmodified versus blocked hyperpolarization-activated cyclic nucleotide gated (HCN) channel\footnote{HCN channels form the biophysical substrate of h-current based adaptation.} recordings to show that in the presence of HCN channels action potential (AP) initiation is enhanced during sustained firing and the propagation of AP patterns is sped up. They observe that in the presence of HCN channels, hyperpolarizing Na$^+$-K$^+$ currents are opposed. \cite{Rich2025} included both HCN (h-current based adaptation) and M-channel (SFA) dynamics in a spiking neural network to investigate their role in seizure events in epilepsy patients. \cite{Porta2024} found, in slice experiments, that blocking h-currents results in prolonged up- and down-state durations, effectively decreasing the temporal oscillation frequency. With their computational model they produce adaptation-driven SOs between up- and down-states where the implemented h-current and SFA based adaptation alternate in activation, depending on the membrane potential with h-currents active in down-, and inactive in up-states while the opposite is the case for SFA. With the model, they recreated the results found in vitro, prolonging up- and down-state durations, if the strength of h-currents is further reduced. \cite{Mehrotra2024} tested whether a spatial gradient in excitability, degree of recurrent excitation, or strength of the SFA current would induce a sequential activation of cortical neurons (i.e, up-state onset) along the dorso-ventral axis of the rat post-subiculum with a synchronous deactivation of the up-state. They observed this in vivo in naturally sleeping mice prior to their in-silico experiments. Neither gradient was able to replicate the experimentally observed dynamics. Including h-currents with decreasing strength along the dorso-ventral axis, on the other hand, reproduced their results. Although previous studies examined each mechanism in detail, a unified investigation comparing both mechanisms, including systematic state-space explorations, has not yet been performed. 

To address this, we implement a Wilson-Cowan field model and consider two nonlinear adaptation mechanisms, a negative feedback mechanism induced by high neural activity and a positive feedback mechanism induced by low neural activity, capturing the most prominent features of SFA- vs h-current based adaptation. The model is implemented on a one-dimensional spatial domain with periodic boundary conditions. Each location contains two interacting populations of excitatory and inhibitory neurons that are spatially coupled by a Gaussian synaptic connectivity kernel. The excitatory populations are locally equipped with either one of the two adaptation mechanisms.  Since both mechanisms act in a functionally similar way, we describe them with the same dynamical equation and consider the inverse mode of action (negative vs.\ positive feedback and activation during high- vs. low-activity states) by changing the sign of the adaptation strength and the gain of the activation function. The model is parameterized to fulfill various criteria that have been shown to play a role in the formation of adaptation-driven slow waves, e.g.,\ regimes of bistability (where stable up- and down-states coexist) in a cortical network for SOs during sleep \cite{Andola2017} and anesthesia \cite{Dasilva2012}, or strong excitatory recurrence \cite{SanchezVivez2000}.

We first show that in our formulation both adaptation mechanisms are mathematically equivalent under a compensatory external input to the excitatory populations that depends on the adaptation strength. A state space analysis shows that strong enough adaptation is necessary to produce spatio-temporal activity patterns in this system that only generates homogeneous or stationary activity when adaptation is weak. 
Secondly, we conducted numerical simulations and show that in the two-population model the adaptation strength is a modulatory factor for the adaptation-driven slow waves. For the considered slices of state space, we observe an increase of dominant temporal frequencies, spatial frequencies, and travel speed of spatio-temporal activity patterns, the stronger the adaptation strength becomes. Numerical results agree with the results of the mathematical analysis performed before.

Due to the dynamical equivalence, the state spaces for the positive and negative feedback mechanisms are equal but shifted along the external input parameter of the excitatory population proportional to the adaptation strength. This finding may serve as a guideline for what differences to expect in a given neurobiological setting (e.g.,\ when analyzing the effects of spatial gradients of adaptation strength as in \cite{Mehrotra2024}). Furthermore, the Wilson-Cowan field model with adaptation can serve as a reference when considering the influence of the additional differences in the physiological properties of SFA- and h-currents-based adaptation mechanisms.

\section{The spatially extended adaptive Wilson-Cowan model}\label{sec:models}
We use the original Wilson-Cowan continuum model of \cite{Wilson1973} with integro-differential equations, including a convolution over a spatial domain, to investigate adaptation-driven traveling waves of SO-like dynamics. We include a third differential equation for an adaptation mechanism and connect it to the excitatory population to describe the effects of either high-activity induced negative or low-activity induced positive feedback mechanisms. In the following we refer to them as spike-frequency ($a$) or h-current ($h$) mediated adaptation although the model is a simplification that focuses only on their most prominent biophysical features. The equations of the activities $u_e$ and $u_i$ of the excitatory and inhibitory population and for the adaptation variable $m$ are given by
\begin{widetext}
\begin{equation}
    \begin{split}
    \tau_e\frac{\partial}{\partial t}u_e(x,t)=&\ -u_e(x,t)+F_e(w_{ee}w_e(x)\ast u_e(x,t)-w_{ei}w_i(x)\ast u_i(x,t)-bm(x,t)+I_e)\\
    \tau_i\frac{\partial}{\partial t}u_i(x,t)=&\ -u_i(x,t)+F_i(w_{ie}w_e(x)\ast u_e(x,t)-w_{ii}w_i(x)\ast u_i(x,t)+I_i)\\
    \tau_{m}\frac{\partial}{\partial t}m(x,t) =&\ -m(x,t)+F_m(u_e(x,t)-\mu),
    \label{eq:model}
\end{split}
\end{equation}\end{widetext}where the spatial convolution is defined as\begin{align*}
    w_j(x)\ast u_j(x,t)=&\ \int_{-\infty}^\infty w_j(|x-y|)\ u_j(y, t)dy,\\
    &\ j\in\{e,i\}.
\end{align*}
The spatial kernel is given by the normalized (i.e.,\ $\int_{-\infty}^\infty w_j(x)dx=1$) Gaussian \begin{align}
    w_j(|x-y|)=\frac{1}{\sqrt{2\pi}\sigma_j}\text{e}^{-\frac{\lVert x-y\rVert^2}{2\sigma_j^2}},
    \label{eq:kernel}
\end{align}with width $\sigma_j,\ j\in\{e,i\}$. The transfer functions are sigmoidal, 
\begin{align}
    F_j(x)=\frac{1}{1+\text{e}^{-\beta_jx}},\ \ j\in\{e,i,m\}.
    \label{eq:tranfer-function}
\end{align}The coupling strengths between the excitatory and inhibitory populations are given by connectivity weights $w_{jl},\ j,l\in\{e,i\}$, in Eq.\ \eqref{eq:model}. The membrane time constants are designated with $\tau_j,\ j\in\{e,i\}$. The adaptation mechanisms $m(x,t)$ are discussed in detail in Section \ref{subsec:neuro-modulators-methods}; they are parametrized by a strength value $b$, a time constant $\tau_m$, and a gain parameter $\beta_m$. Due to the transfer function in Eq.\ \eqref{eq:tranfer-function} being a sigmoid between zero and one, we interpret the activity variables $u_e$ and $u_i$ as the proportion of neurons firing per unit time \cite{Kilpatrick2013WC}. Its gain parameter is denoted by $\beta_j$. The kernel widths in Eq.\ \eqref{eq:kernel} determine the extent to which populations are effectively connected in space. Kernel widths are chosen to be smaller for excitatory and broader for inhibitory interactions, effectively implementing a Mexican-hat shaped interaction between different locations. The parameters that are used in this study were adjusted such that adaptation-driven SOs are generated with an additional noisy external input (see \cite{Cakan2022, Dimulescu2025}), fast excitation-inhibition driven oscillations can emerge (see \cite{Papadopoulos2020}), and the system with and without adaptation shows regimes of bistability since this has been shown to play a crucial part in the formation of SOs (see \cite{Andola2017, Sanchez-Vives2020}).\\
\begin{table}[h]
    \centering
    \caption{List of model parameters. The abbreviations ``exc'' and ``inh'' stand for ``excitatory population'' and ``inhibitory population'', respectively.}
    \begin{tabular}{|l|c|r|}
        \hline
        Parameter & Value & Description \\
        \hline
        \hline
        $\mu_e$ & 0 & Excitatory threshold \\
        $\mu_i$ & 0 & Inhibitory threshold \\
        $\beta_e$ & 5 & Excitatory gain \\
        $\beta_i$ & 5 & Inhibitory gain \\
        \hline
        \hline
        $w_{ee}$ & 3.2 & Coupling strength exc to exc \\
        $w_{ei}$ & 2.6 & Coupling strength inh to exc \\
        $w_{ie}$ & 3.3 & Coupling strength exc to inh \\
        $w_{ii}$ & 0.9 & Coupling strength inh to inh \\
        $I_e$ & [-2.5, 2] & External input current to exc \\ 
        $I_i$ & [-3.5, 1] & External input current to inh \\ 
        $\tau_e$ & $10$ & Membrane time constant of exc \\ 
        $\tau_i$ & $15$ & Membrane time constant of inh \\ 
        $\sigma_e$ & 1 & Spatial spread of exc \\ 
        $\sigma_i$ & 3 & Spatial spread of inh\\
        $\mu$ & $0.4$ & Adaptation threshold \\
        \hline 
        \hline 
        $dx$ & $0.3$ & Integration space step \\ 
        $L$ & $600$ & Length (circumference) of ring\\
        \hline
        \hline
        $b$ & $[0,1]$ & SFA strength \\
        $\tau_m$ & $300$ & SFA time constant \\
        $\beta_m$ & $10$ & SFA gain \\
        \hline
        \hline
        $-b$ & $[0,1]$ & H-current strength \\
        $\tau_m$ & $300$ & H-current time constant \\
        $-\beta_m$ & $10$ & H-current gain \\
        \hline
    \end{tabular}
    \label{tab:default-params}
\end{table}
The model is implemented as a one-dimensional neural field with periodic boundary conditions and a homogeneous connectivity profile. To reduce the effects of the finite size on the dynamics, we combine a large ring of  length $L=600$ with comparably small widths $\sigma_j,\ j\in\{e,i\}$, of the spatial kernels. The ring is discretized into $n$ positions, each of which is equipped with a node whose dynamics is described by Eqs.\ \eqref{eq:model}. Therefore, the integration space step is given by $dx=\frac{L}{n}$.
\subsection{The adaptation mechanisms}\label{subsec:neuro-modulators-methods}
The excitatory populations are equipped with an adaptation mechanism, generally understood as a process (not restricted to the brain) that modulates the effect of a sustained input but acts on a slower time scale (see \cite{Friedlander2009}). In the following, we distinguish two adaptation processes caused by different types of somatic conductances in cortical neurons (see \cite{Timofeev2002}). Spike-frequency adaptation (SFA), a negative feedback mechanism, is mediated by slow potassium currents which activate when the activity of the excitatory population is high. This causes a hyperpolarizing feedback current which effectively inhibits the neural response of excitatory neurons (see \cite{Ladenbauer2014}). H-current based adaptation is mediated by hyperpolarization-activated nucleotide-gated channels (see \cite{Hong-Yuan2010}) which open if the excitatory population is in a low-activity state and return positive feedback, i.e.,\ effectively excite the excitatory population. Functionally, the adaptation mechanisms are similar (see \cite{Ganguly2024} for SFA, for h-currents see \cite{McCormick1990} where the authors discuss h-currents in thalamic neurons, and see \cite{Mehrotra2024, Timofeev2002} for both adaptation mechanisms in cortical neurons), so that we can describe their main effects by the same differential equation (last equation in Eqs.\ \eqref{eq:model}, see \cite{Mehrotra2024}) and include the inverse mode of action (positive versus negative feedback, activated during high- versus low-activity values) by changing the sign of the parameters for adaptation strength $b$ and adaptation gain $\beta_m$. Parameters are summarized in Table \ref{tab:default-params}. 
\subsection{Fixed points and stability analysis of the model with adaptation}\label{sec:3d-stability-analysis-methods}
To determine the fixed points we consider a single Wilson-Cowan node. Due to the normalization of the kernel, its fixed points are equivalent to the fixed points of the spatially extended model. Let $\frac{\partial u_e}{\partial t} = \frac{\partial u_i}{\partial t} = 0 $ in Eqs.\ \eqref{eq:model}. From $\frac{\partial m}{\partial t}=0$ we obtain $m=F_m(u_e-\mu)$ which we insert in the equation for the excitatory population. The nullclines $f(u_e)$ and $g(u_i)$ are then given by 
\begin{align*}
    u_i = f(u_e) =& \frac{1}{w_{ei}}\bigl(w_{ee}u_e + I_e -bF_m(u_e-\mu) - F^{-1}_e(u_e)\ \bigr),\\
    u_e = g(u_i) =& \frac{1}{w_{ie}}\bigl(w_{ii}u_i - I_i + F^{-1}_i(u_i)\ \bigr).
\end{align*}A point $(\tilde{u}_e, \tilde{u}_i, \tilde{m})\in(0,1)^3$ is a fixed point if $f(\tilde{u}_e)=\tilde{u}_i$, $g(\tilde{u}_i)=\tilde{u}_e$, and $F_m(\tilde{u}_e-\mu)=\tilde{m}$. 

We apply the procedure from Wyller et al \cite{Wyller2007}\footnote{\cite{Wyller2007}, however, assumes that the fixed point values $\tilde{u}_e$ and $\tilde{u}_i$ are equal.}, Harris \& Ermentrout \cite{Harris2018}\footnote{\cite{Harris2018} investigate the stability of homogeneous equilibria for the two-population model (i.e., $b=0$).}, and Byrne et al \cite{Byrne2019}  to investigate the stability of the homogeneous steady states (i.e.,\ the fixed points) of field models. To do so, we add a small spatial perturbation to the homogeneous steady state,
\begin{align}
\begin{split}
    u_e(x,t)&=\tilde{u}_e+\chi(x,t)\\
    u_i(x,t)&=\tilde{u}_i +\varphi(x,t)\\
    m(x,t)&=\tilde{m} +\phi(x,t).
\end{split}
    \label{eq:perturbation}
\end{align}
\begin{figure*}[t]
    \centering
    \includegraphics[width=0.99\linewidth]{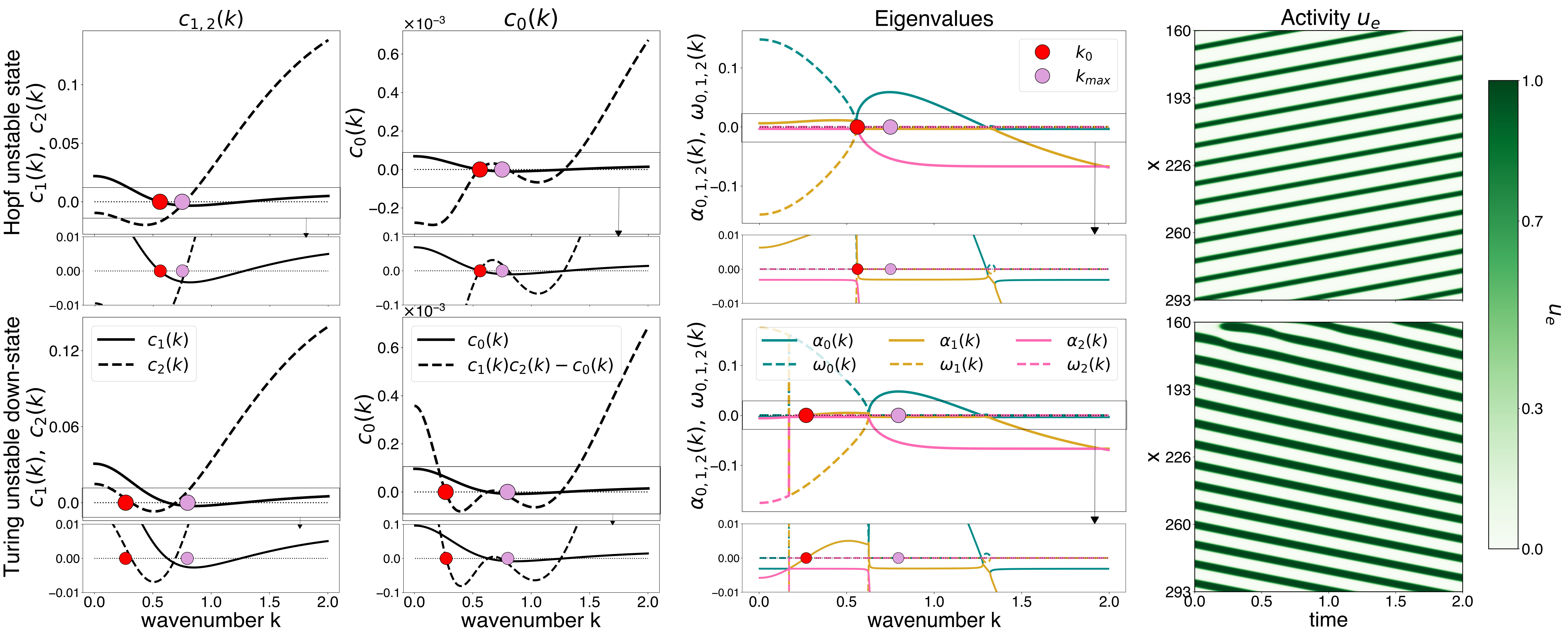}
    \caption{Examples for static (top row) and dynamic (bottom row) Turing instabilities  in a Hopf- (top row) and Turing-unstable state (bottom row) for the system with h-currents. Panels from left to right show (cf. Eqs.\ \eqref{eq:cs}): Coefficients $c_{1}(k)$ (solid line) and $c_2(k)$ (dashed line), coefficient $c_0(k)$ (solid line) and condition $c_1(k)c_2(k)-c_0(k)$ (dashed line), real ($\alpha_{0,1,2}(k)$, solid lines) and imaginary ($\omega_{0,1,2}(k)$, dashed lines) parts of the eigenvalues of the linearization matrix, and the corresponding activity pattern $u_e(x,t)$. Circles denote the wavenumbers $k_0$ (red) and $k_{max}$ (pink) (see text). Insets below each plot show close-ups of the thin boxes. Values for the external input currents and the adaptation parameter were $(I_e, I_i) = (-0.7, -0.6),\ b=-0.5$ (top row) and $(I_e, I_i) = (-0.3, -0.3),\ b=-0.5$ (bottom row). All other parameters are given in Table \ref{tab:default-params}.}
    \label{fig:wavenumbers}
\end{figure*}
Then, we insert Eqs.\ \eqref{eq:perturbation} into Eqs.\ \eqref{eq:model} and linearize the system around the homogeneous steady state by applying a Taylor expansion up to the first order in $\chi(x,t)$, $\varphi(x,t)$, and $\phi(x,t)$. We obtain
\begin{widetext}
    \begin{align*}
    \tau_e\partial_t\chi &= -\chi + w_{ee}F^\prime_e(B_e)w_e\ast\chi - w_{ei}F^\prime_e(B_e)w_i\ast\varphi -bF_e^\prime(B_e)\phi \\
    \tau_i\partial_t\varphi &= -\varphi + w_{ie}F^\prime_i(B_i)w_e\ast\chi - w_{ii}F^\prime_i(B_i)w_i\ast\varphi \\
    \tau_m\partial_t\phi &= -\phi + F^\prime_m(\tilde{u}_e-\mu)\chi,
\end{align*}
\end{widetext}where $B_e:=w_{ee}\tilde{u}_e-w_{ei}\tilde{u}_i-bF_m(\tilde{u}_e-\mu)+I_e$ and $B_i:=w_{ie}\tilde{u}_e-w_{ii}\tilde{u}_i+I_i$. The zero order terms vanish for $\tilde{u}_e$ and $\tilde{u}_i$, because the spatial kernels are normalized to one. To transform this set of IDEs into a set of ordinary differential equations (ODEs), we exploit the fact that a convolution in the spatial domain is a product in the frequency domain (i.e.,\ $w(x)\ast\chi(x,t)=\hat{w}(k)\hat{\chi}(k,t)$, where $\hat{\cdot}$ denotes the Fourier Transformation). This leads to the ODE system
\begin{widetext}\begin{align}
    \begin{split}
    \tau_e\partial_t\hat{\chi} &= -\hat{\chi} + w_{ee}F^\prime_e(B_e)\hat{w}_e\hat{\chi} - w_{ei}F^\prime_e(B_e)\hat{w}_i\hat{\varphi} -bF_e^\prime(B_e)\hat{\phi}\\
    \tau_i\partial_t\hat{\varphi} &= -\hat{\varphi} + w_{ie}F^\prime_i(B_i)\hat{w}_e\hat{\chi} - w_{ii}F^\prime_i(B_i)\hat{w}_i\hat{\varphi}\\
    \tau_m\partial_t\hat{\phi} &= -\hat{\phi} + F^\prime_m(\tilde{u}_e-\mu)\hat{\chi}.
    \end{split}
    \label{eq:space-fourier-system}
\end{align}\end{widetext}For $\vec{X}=\begin{pmatrix} \hat{\chi} \\ \hat{\varphi} \\ \hat{\phi} \end{pmatrix}$, Eqs.\ \eqref{eq:space-fourier-system} can be written in vector notation, $\partial_t \vec{X}=A_3(k)\vec{X}$, where
\begin{widetext}\begin{align}
\begin{split}
    A_3(k)=\begin{bmatrix}
        -\frac{1}{\tau_e}+\frac{w_{ee}}{\tau_e}F^\prime_e(B_e)\hat{w}_e(k) & -\frac{w_{ei}}{\tau_e}F^\prime_e(B_e)\hat{w}_i(k)& -\frac{b}{\tau_e}F^\prime_e(B_e) \\
        \frac{w_{ie}}{\tau_i}F^\prime_i(B_i)\hat{w}_e(k) & -\frac{1}{\tau_i}-\frac{w_{ii}}{\tau_i}F^\prime_i(B_i)\hat{w}_i(k) & 0 \\
        \frac{F^\prime_m(\tilde{u}_e)}{\tau_m} & 0 & -\frac{1}{\tau_m}
    \end{bmatrix}
        =: \begin{bmatrix}
        a_{11}(k) & a_{12}(k) & a_{13} \\
        a_{21}(k) & a_{22}(k) & a_{23} \\
        a_{31} & a_{32} & a_{33}
    \end{bmatrix},
\end{split}
    \label{eq:spatial-adaps-jacobian}
\end{align}\end{widetext}$B_e:=w_{ee}\tilde{u}_e-w_{ei}\tilde{u}_i-bF_m(\tilde{u}_e-\mu)+I_e$ and $B_i:=w_{ie}\tilde{u}_e-w_{ii}\tilde{u}_i+I_i$. The stability of the fixed point can be characterized by the eigenvalues $\lambda_{j}(k)=\alpha_j(k)+i\omega_j(k),\ j\in\{0,1,2\}$, of the linearization matrix (Jacobian) $A_3(k)$, where $i$ denotes the imaginary unit. The eigenvalues correspond to the roots of the cubic polynomial
\begin{widetext}
\begin{align*}
    \det(A_3(k)-\lambda\underline{\text{I}})=&\ \lambda^3+\lambda^2\bigl(-(a_{11}(k)+a_{22}(k)+a_{33})\bigr) \notag\\
    & +\lambda\bigl(a_{11}(k)a_{22}(k)+a_{11}(k)a_{33}+a_{22}(k)a_{33}-a_{13}a_{31}-a_{12}(k)a_{21}(k)\bigr) \notag \\
    & + \bigl(-a_{11}(k)a_{22}(k)a_{33}+a_{22}(k)a_{13}a_{31}+a_{33}a_{12}(k)a_{21}(k)\bigr) \notag \\
     =&\ \lambda^3+\lambda^2 c_2(k)+\lambda c_1(k)+c_0(k)\label{eq:3dPoly}
\end{align*}
\end{widetext}
with
\begin{equation}
    \begin{split}
        c_0(k) =&\ -a_{11}(k)a_{22}(k)a_{33}+a_{22}(k)a_{13}a_{31}\\
        &\ +a_{33}a_{12}(k)a_{21}(k)\\
        c_1(k) =&\ a_{11}(k)a_{22}(k)+a_{11}(k)a_{33}+a_{22}(k)a_{33}\\
        &\ -a_{13}a_{31}-a_{12}(k)a_{21}(k)\\
        c_2(k) =&\ -(a_{11}(k)+a_{22}(k)+a_{33}).
    \end{split}
    \label{eq:cs}
\end{equation}
Fixed points with $\alpha_j(k)<0$ for all $j\in\{0,1,2\}$ are stable. 


\subsubsection{Instabilities of homogeneous equilibria}
In the following, we investigate the stability of homogeneous equilibria in analogy to \cite{Wyller2007}. If the model system, for a fixed parameterization, has a fixed point which is unstable at $k=0$ (i.e.,\ $\alpha_j(0)\geq0$ for at least one $j\in\{0,1,2\}$, see Fig.\ \ref{fig:wavenumbers}, first row), we call the fixed point \textit{Hopf-unstable}, if $\omega_j(0)\neq0$, or simply \textit{unstable}, if $\omega_j(0)=0$. 

A fixed point $(\tilde{u}_e, \tilde{u}_i, \tilde{m})$ corresponds to a generalized Turing situation at $k_0>0$ if $c_0(k_0)=0$, i.e.,\ if one of the eigenvalues is zero for this wavenumber. This is illustrated in Fig.\ \ref{fig:wavenumbers}, first row, where the condition is met for the wavenumber $k_0$ indicated by the red circle. The instability occurring at $k_0$ can also be called static \cite{Meijer2014}. If $c_2(k_0)c_1(k_0)-c_0(k_0)=0$ for $c_1(k_0)>0$, i.e.,\ if one eigenvalue at this wavenumber is purely imaginary, the fixed point corresponds to a finite bandwidth instability with oscillations (also called dynamic \cite{Meijer2014} or a Turing-Hopf instability, see \cite{Byrne2019}). This is illustrated in Fig.\ \ref{fig:wavenumbers}, second row, where the condition is met, for example, for the wavenumber indicated by the red marker. Since all bands of unstable modes are of finite width (see Appendix \ref{appendix:finite-bandwidthW}), at least two zero-crossing occur for some $k_1>k_0>0$ if the fixed point displays a Turing instability. If both types (i.e.,\ static and dynamic) of Turing situations occur, it is called a mixed Turing situation. We refer to a fixed point that is stable at $k=0$ for which a a static, dynamic or mixed Turing situation occurs, under the umbrella term ``\textit{Turing-unstable}''. For the investigated parameterizations, Hopf-unstable states may also have bands of unstable modes at non-zero wavenumbers $k>0$ (see, e.g.,\  Fig.\ \ref{fig:wavenumbers}, first row). 

Since we are particularly interested in adaptation-driven slow alternations between high- (up) and low-activity (down) states, we distinguish up- ($\tilde{u}_e \geq 0.4$) from down- ($\tilde{u}_e < 0.4$) states by the fixed point value of the excitatory population. Consequently, when a Turing instability emerges in an up- (down-) state, we call the state \textit{Turing-unstable up- (down-) state}. In addition, the model exhibits regimes of bistability between two fixed points where stable up- and down-states coexist for the same location in state space. If either one of the two states loses stability for some $k_0>0$ due to an emerging Turing situation while the coexisting fixed point remains stable, we call the Turing-unstable state a \textit{Turing unstable up- or down-state in bi}, respectively. 

Note that there is a difference in the instability of modes between Hopf-unstable states with bands of unstable modes with non-zero wavenumbers $k>0$ and other Turing-unstable states which is exemplified in Fig.\ \ref{fig:wavenumbers}, third column (Eigenvalues). In Hopf-unstable states, unstable modes already occur for wavenumber values $0\leq k \leq k_0$, while in Turing-unstable states the minimum wavenumber for unstable modes is $k_0>0$. 
\subsection{Wavenumbers}\label{subsec:wavenumber}
If a fixed point is identified as Turing- or Hopf-unstable, we compute the smallest wavenumber $k_0>0$ at which the conditions for one of the above-mentioned Turing instabilities are fulfilled. Examples are shown in Fig.\ \ref{fig:wavenumbers}, where $k_0$  is marked by the red circles. Additionally, we determine the wavenumber $k_{max}$ (see pink circles in Fig.\ \ref{fig:wavenumbers}) for which the corresponding unstable mode grows most rapidly, i.e.,\ for which the real part $\alpha_j(k)$ of the largest eigenvalue becomes maximal, $k_{max}=\argmax_k(\max_j\alpha_j(k))$. 

\subsection{Symmetries and equivalence of mechanisms}\label{sec:equivalency-transformation}
Let Eqs.\ \eqref{eq:model},
\begin{widetext}{\small{\begin{align}
    S(u_e, u_i, m)=\begin{pmatrix} \tau_e\frac{\partial u_e}{\partial t} \\ \tau_i\frac{\partial u_i}{\partial t} \\ \tau_m\frac{\partial m}{\partial t}\end{pmatrix} = \begin{pmatrix}
        -u_e(x,t)+F_e(w_{ee}w_e(x)\ast u_e(x,t)-w_{ei}w_i(x)\ast u_i(x,t)-bm(x,t)+I_e) \\
        -u_i(x,t)+F_i(w_{ie}w_e(x)\ast u_e(x,t)-w_{ii}w_i(x)\ast u_i(x,t)+I_i)\\
        -m(x,t)+F_m(u_e(x,t)-\mu)
    \end{pmatrix},\label{eq:matrix-model}
\end{align}}}\end{widetext}be the system with SFA (i.e.,\ let $b,\ \beta_m > 0$), and let $T:[0,1]^3\to[0,1]^3$ with \begin{align*}T: (u_e, u_i, m)^\top \mapsto (u_e, u_i, 1-m)^\top\end{align*} be an isomorphic transformation. Then we have $\frac{\partial}{\partial t}\bigl[ T\bigl((u_e(x,t), u_i(x,t), m(x,t))^\top\bigr)\bigr] = (\frac{\partial  u_e}{\partial t},  \frac{\partial  u_i}{\partial t}, -\frac{\partial  m}{\partial t})^\top$. Inserting the transformation into Eqs.\ \eqref{eq:matrix-model} we obtain \begin{widetext}{\small{\begin{align*}
    S(\mathbf{T(u_e, u_i, m)})=\begin{pmatrix} \tau_e\frac{\partial u_e}{\partial t} \\ \tau_i\frac{\partial u_i}{\partial t} \\ \mathbf{-\tau_m\frac{\partial m}{\partial t}}\end{pmatrix} =& \begin{pmatrix}
        -u_e(x,t)+F_e(w_{ee}w_e(x)\ast u_e(x,t)-w_{ei}w_i(x)\ast u_i(x,t)-\mathbf{b(1-m(x,t))}+I_e) \\
        -u_i(x,t)+F_i(w_{ie}w_e(x)\ast u_e(x,t)-w_{ii}w_i(x)\ast u_i(x,t)+I_i)\\
        -\bigl(\mathbf{-(1-m(x,t)})+\underbrace{F_m(u_e(x,t)-\mu)}_{=1-F_m(\mu-u_e)}\bigr)\end{pmatrix}\notag \\
        =& 
        \begin{pmatrix}-u_e(x,t)+F_e(w_{ee}w_e(x)\ast u_e(x,t)-w_{ei}w_i(x)\ast u_i(x,t) + bm(x,t) -b+I_e) \\
        -u_i(x,t)+F_i(w_{ie}w_e(x)\ast u_e(x,t)-w_{ii}w_i(x)\ast u_i(x,t)+I_i)\\
        -m(x,t)+F_m(\mu-u_e(x,t))
    \end{pmatrix}.
\end{align*}}}\end{widetext}

\begin{figure*}[t]
    \centering
    \includegraphics[width=0.99\linewidth]{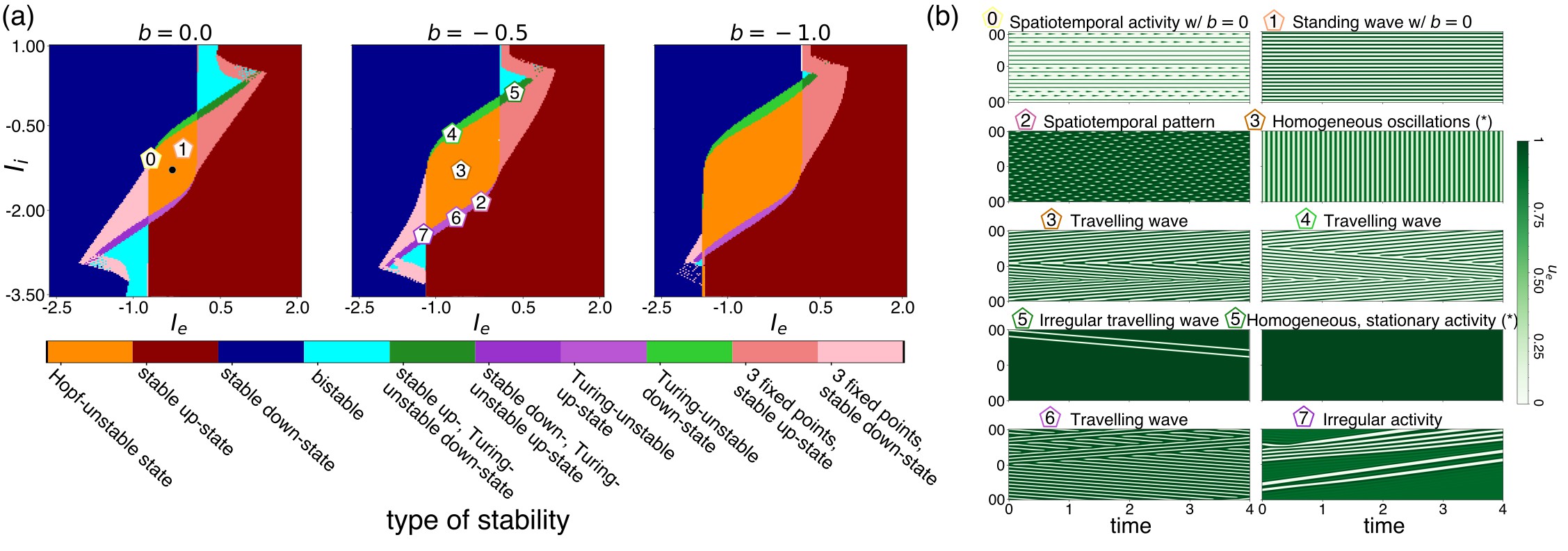}
    \caption{State spaces and activity patterns for the Wilson-Cowan model with h-currents. \textbf{(a)} Results of the stability analyses for a slice of parameter space spanned by the external input currents $(I_e, I_i)\in[-2.5,2]\times[-3.5,1]$ for increasing adaptation strength $|b|\in\{0, 0.5, 1.0\}$ (left to right). Colors denote the different dynamical regimes (see legend). Pentagon markers identify the positions in state space corresponding to the activity traces shown in \rm{(b)}. Resolution of state space is \texttt{181 x 181}. The black dot in the state space for $b=0$ denotes center of point symmetry (see Section \ref{sec:equivalency-transformation}). \textbf{(b)} Activity traces $u_e(x,t)$ for different locations in state space (see Table \ref{tab:state-space-input-pairs} in Appendix \ref{appendix:state-space-activity-pairs}), denoted by pentagons in \rm{(a)}. Two types of bistability are exemplified in panels 3, 3(*) and 5, 5(*) (see text). Lighter green denotes low, darker green high activity values (see color bar). All simulations were initialized around their corresponding fixed points, except for the activity traces marked with (*) which were initialized close to zero (see Appendix \ref{subsec:initialisation-methods}).}
    \label{fig:equivalency-results-with-activity}
\end{figure*}

This system of equations is, given the choice of $F_m$ (Eq.\ \eqref{eq:tranfer-function}), equivalent to the system of Eqs.\ \eqref{eq:model} for h-currents if $-b\leftarrow b$ and $-\beta_m\leftarrow \beta_m$ (see Table \ref{tab:default-params}) and the external input to the excitatory population in the system with SFA is decreased by the value of $b$, i.e.,\  $I_e^h=I_e^a-b$. The superscripts $\cdot^a$, $\cdot^h$ indicate the difference between the variables and parameters for SFA and h-currents, respectively. Hence, both adaptation mechanisms are dynamically equivalent under the transformation $T$, if $I_e$ is correspondingly changed and if $F_m^a(x)=1-F_m^h(x)$ holds for the adaptation nonlinearity. This equivalence also tells us that a fixed point $(\tilde{u}_e^a, \tilde{u}_i^a, \tilde{m}^a)$ for the system with SFA corresponds to a fixed point $(\tilde{u}_e^h, \tilde{u}_i^h, 1-\tilde{m}^h)$ for the system with h-currents and that the dynamical landscape for h-currents should be the same as the dynamical landscape for SFA, but shifted along the $I_e$-axis by the value $b$.

Beyond dynamical equivalence between the adaptation mechanisms, we also have a symmetry for each system within itself. Let $T:[0,1]^3\to[0,1]^3$ with $T: (u_e, u_i, m)^\top \mapsto (1-u_e, 1-u_i, 1-m)^\top$ be another isomorphic transformation such that $\frac{\partial}{\partial t}\bigl[ T\bigl((u_e(x,t), u_i(x,t), m(x,t))^\top\bigr)\bigr] = (-\frac{\partial  u_e}{\partial t},  -\frac{\partial  u_i}{\partial t}, -\frac{\partial  m}{\partial t})^\top$. Inserting this in Eqs.\ \eqref{eq:matrix-model} we obtain \begin{widetext}{\small\begin{align}
    \begin{pmatrix} -\tau_e\frac{\partial u_e}{\partial t} \\ -\tau_i\frac{\partial u_i}{\partial t} \\ -\tau_m\frac{\partial m}{\partial t}\end{pmatrix} =& 
   \begin{pmatrix}
        1 - u_e(x,t)-F_e(w_{ee}w_e(x)\ast (1-u_e(x,t))-w_{ei}w_i(x)\ast (1-u_i(x,t))-b(1-m(x,t))+I_e) \\
        1 - u_i(x,t)-F_i(w_{ie}w_e(x)\ast (1-u_e(x,t))-w_{ii}w_i(x)\ast (1-u_i(x,t))+I_i)\\
        1 - m(x,t)-F_m((1-u_e(x,t))-\mu)\end{pmatrix} \notag \\
         =&
        \begin{pmatrix}
        -u_e(x,t)+F_e(w_{ee}w_e(x)\ast u_e(x,t)-w_{ei}w_i(x)\ast u_i(x,t) - bm(x,t) - I_e -w_{ee} + w_{ei} + b) \\
        -u_i(x,t)+F_i(w_{ie}w_e(x)\ast u_e(x,t)-w_{ii}w_i(x)\ast u_i(x,t) - I_i - w_{ie} + w_{ii})\\
        -m(x,t)+F_m(u_e(x,t) + \mu -1)
    \end{pmatrix}. \label{eq:point-symmetry}
\end{align}}\end{widetext}If $\hat{I}_e=-I_e+b-w_{ee}+w_{ei},\ \hat{I}_i=-I_i-w_{ie}+w_{ii}$, and $\hat{\mu}=1-\mu$, then the system \eqref{eq:point-symmetry} is equivalent to Eqs.\ \eqref{eq:model} for both SFA and h-currents. This induces an approximate point symmetry of state space at $(I^c_e, I^c_i)=\left(\frac{b-w_{ee}+w_{ei}}{2} , \frac{-w_{ie}+w_{ii}}{2}\right)$ between $(u_e, u_i, m)$ and $(1-u_e, 1-u_i, 1-m)$ which becomes exact for $\mu=0.5$. Since $\mu=0.4$ in this study, an exact symmetry is only achieved for $b=0$ (Fig.\ \crefformat{figure}{#2#1{(a)}#3}\cref{fig:equivalency-results-with-activity}, first column, where the black dot in the middle of the orange regime denotes the center of the point reflection).
\section{Overview of state space and its changes with adaptation strength}\label{sec:state-spaces}
Figure \ref{fig:equivalency-results-with-activity} shows the results of the analytical investigation for the system with h-currents with increasing adaptation strength (Fig.\ \crefformat{figure}{#2#1{(a)}#3}\cref{fig:equivalency-results-with-activity}) and examples for the emerging spatio-temporal patterns (Fig.\ \crefformat{figure}{#2#1{(b)}#3}\cref{fig:equivalency-results-with-activity}) for all regimes of Hopf or Turing instability. Without adaptation ($b=0$), we observe a Hopf-unstable regime (orange) in which both temporal (similar to Fig.\ \crefformat{figure}{#2#1{(b)}#3}\cref{fig:equivalency-results-with-activity}, panel 3${(*)}$, for $b=-0.5$) and spatial (see Fig.\ \crefformat{figure}{#2#1{(b)}#3}\cref{fig:equivalency-results-with-activity}, panel 1) patterns emerge. With adaptation, the Hopf regime is also apparent but the activity patterns can be either temporal (Fig.\ \crefformat{figure}{#2#1{(b)}#3}\cref{fig:equivalency-results-with-activity}, panel 3$(*)$) or spatio-temporal (Fig.\ \crefformat{figure}{#2#1{(b)}#3}\cref{fig:equivalency-results-with-activity}, panel 3). Close to the boundaries of the Hopf-unstable regime, we observe the emergence of Turing instability in down-states (upper border; for an activity example see Fig.\ \crefformat{figure}{#2#1{(b)}#3}\cref{fig:equivalency-results-with-activity}, panel 4) and in up-states (lower border; for activity example see Fig.\ \crefformat{figure}{#2#1{(b)}#3}\cref{fig:equivalency-results-with-activity}, panel 6). Beyond the Turing instability in down- (up-) states, to the left (right) side of the state spaces in Fig.\ \crefformat{figure}{#2#1{(a)}#3}\cref{fig:equivalency-results-with-activity}, there exists a regime, marked dark blue (dark red), where a single down- (up-) state fixed point is stable. We also see regimes of three fixed points including one stable up-state (dark beige) while the other two fixed points are unstable and vice versa with one stable down-state (light beige). Activity traces in these regimes are similar to the activities shown in Fig.\ \crefformat{figure}{#2#1{(b)}#3}\cref{fig:equivalency-results-with-activity}, panels 5(*) and 7. Initializations close to zero always converge to the existing stable fixed point while initializations around the unstable states cause irregular traveling waves. Further, to both corner ends (upper right and lower left), we find a regime of bistability between up- and down-states (light blue) next to regimes, where a stable down-state coexists with a Turing instability in the up-state (dark purple) and vice versa (dark green). Figure \crefformat{figure}{#2#1{(b)}#3}\cref{fig:equivalency-results-with-activity}, panel 5, shows that the activity converges to the activity caused by the Turing instability in the down-state, if initialized close to the Turing-unstable down-state. Initialized close to zero, the activity converges to the stable up-state (see activity 5${(*)}$). The pattern emerging in the Turing-unstable down-state of panel 5 shows down-state waves traveling over an otherwise stable up-state activity, unlike the traveling waves shown in panels 4 and 6, where no other stable fixed point coexists. Furthermore, in some states, the activity patterns can become irregular (see Fig.\ \crefformat{figure}{#2#1{(b)}#3}\cref{fig:equivalency-results-with-activity}, panel 7), such that they do not converge to a periodic, equidistant, or otherwise regular pattern. Such irregular patterns predominantly occur at the border of different regimes or in states where another attractor, e.g.,\ a stable fixed point, coexists.

The Hopf-unstable states further experience spatial instability for some $k>0$ (see, for example, Fig.\ \ref{fig:wavenumbers}, top row, third panel), giving rise to spatial (for $b=0$, static Turing instability) or spatio-temporal (for $b\neq0$, dynamic Turing instability) patterns. Initialized close to the fixed points, states in the Hopf regime converge to spatial or spatio-temporal patterns, while other initializations, e.g. initializations close to zero, converge to  fast, spatially homogeneous, temporal oscillations (see also Appendix \ref{appendix:robustness}).
In this regime, the system thus experiences a second type of bi- (or potentially multi-) stability that is not caused by the coexistence of multiple fixed points. 
However, there are two small areas in the Hopf-unstable regime for $b=0$ where exceptions occur and where spatio-temporal patterns are present (Fig.\ \crefformat{figure}{#2#1{(b)}#3}\cref{fig:equivalency-results-with-activity}, panel 0). These areas are located in the top left of the Hopf regime where a stable down-state, three fixed points with a stable down-state, a Hopf instability, and a Turing instability are getting close in parameter space, and in the bottom right corner, where the regimes of a stable up-state, a Hopf instability, three fixed points with a stable up-state, and a Turing instability meet. In these states, depending on the initialization, most of the ring converges toward the standing wave interspersed by a few traveling bumps (see Fig.\ \crefformat{figure}{#2#1{(b)}#3}\cref{fig:equivalency-results-with-activity}, panel 0). When adaptation strength increases, the regimes of bistability between two stable fixed points (light blue) shrink and vanish for values $|b|>1$. The size of the Hopf-unstable regime (orange) and of the regime of three fixed points with a stable up-state (beige) increase, while the regime of three fixed points with a stable down-state decreases in size (light beige).
\begin{figure}[t]
    \centering
    \includegraphics[width=0.99\linewidth]{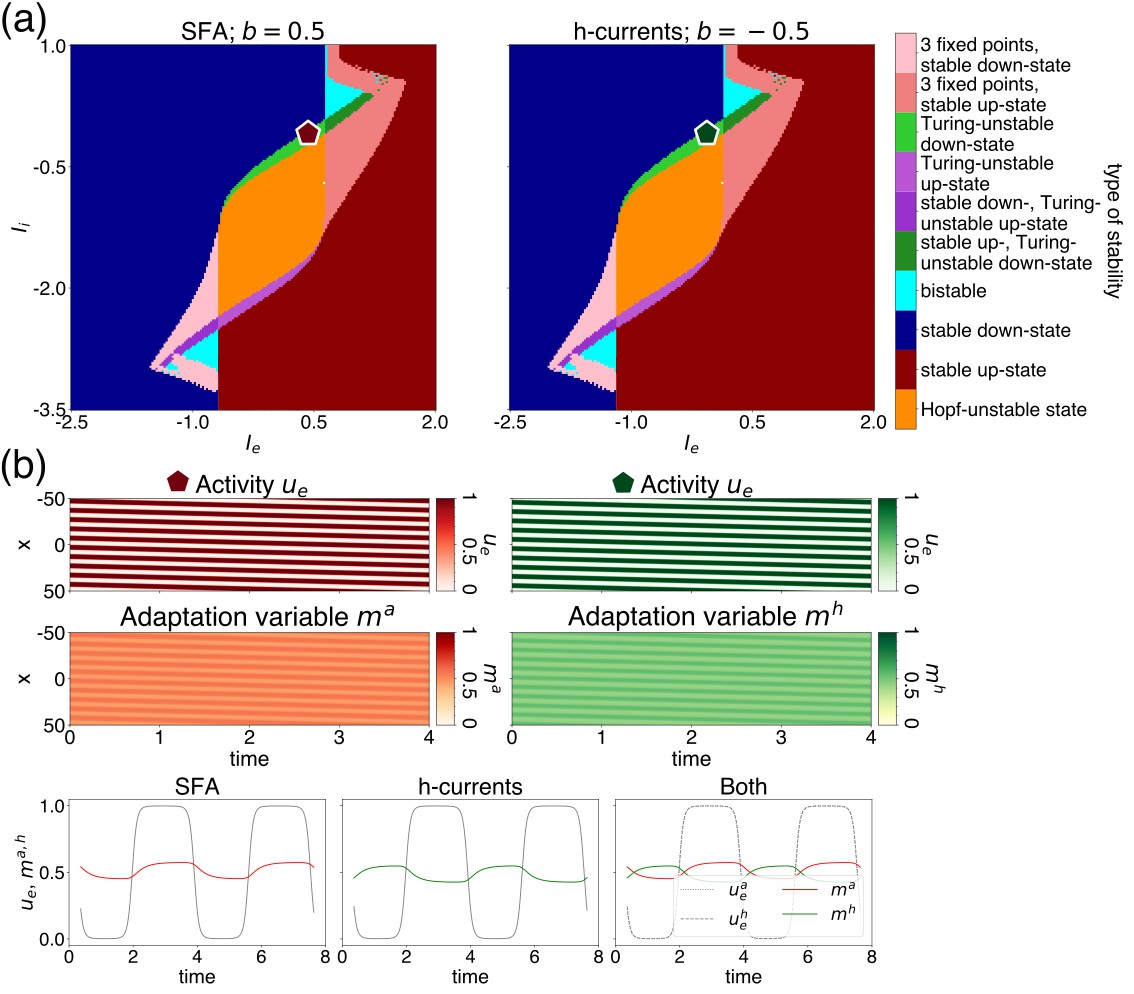}
    \caption{Comparison of state spaces and activity patterns for both adaptation mechanisms. \textbf{(a)} Slices of state space for SFA (left) and h-currents (right) for $|b|=0.5$ spanned by the external input currents $I_e$ and $I_i$. Colors denote the different dynamical regimes (see legend). Colored pentagon markers identify the positions in state space corresponding to the activity traces shown in \rm{(b)}. The external input currents are $(I_e, I_i)=(0.4, -0.3)$ for SFA and $(I_e, I_i)=(-0.1, -0.3)$ for h-currents. \textbf{(b)} Top row: Activity traces $u_e(x,t)$ for the locations in state space, marked in \rm{(a)}. Middle row: Corresponding traces of the adaptation variable $m$. Bottom row: Activity traces and traces of the corresponding adaptation variable for a fixed spatial location for the traveling waves shown above for SFA (left panel) and h-currents (middle panel). All traces are superimposed in the right panel.}
    \label{fig:equivalency-results-with-traces}
\end{figure}
\begin{figure*}[t]
    \centering
    \includegraphics[width=0.99\textwidth]{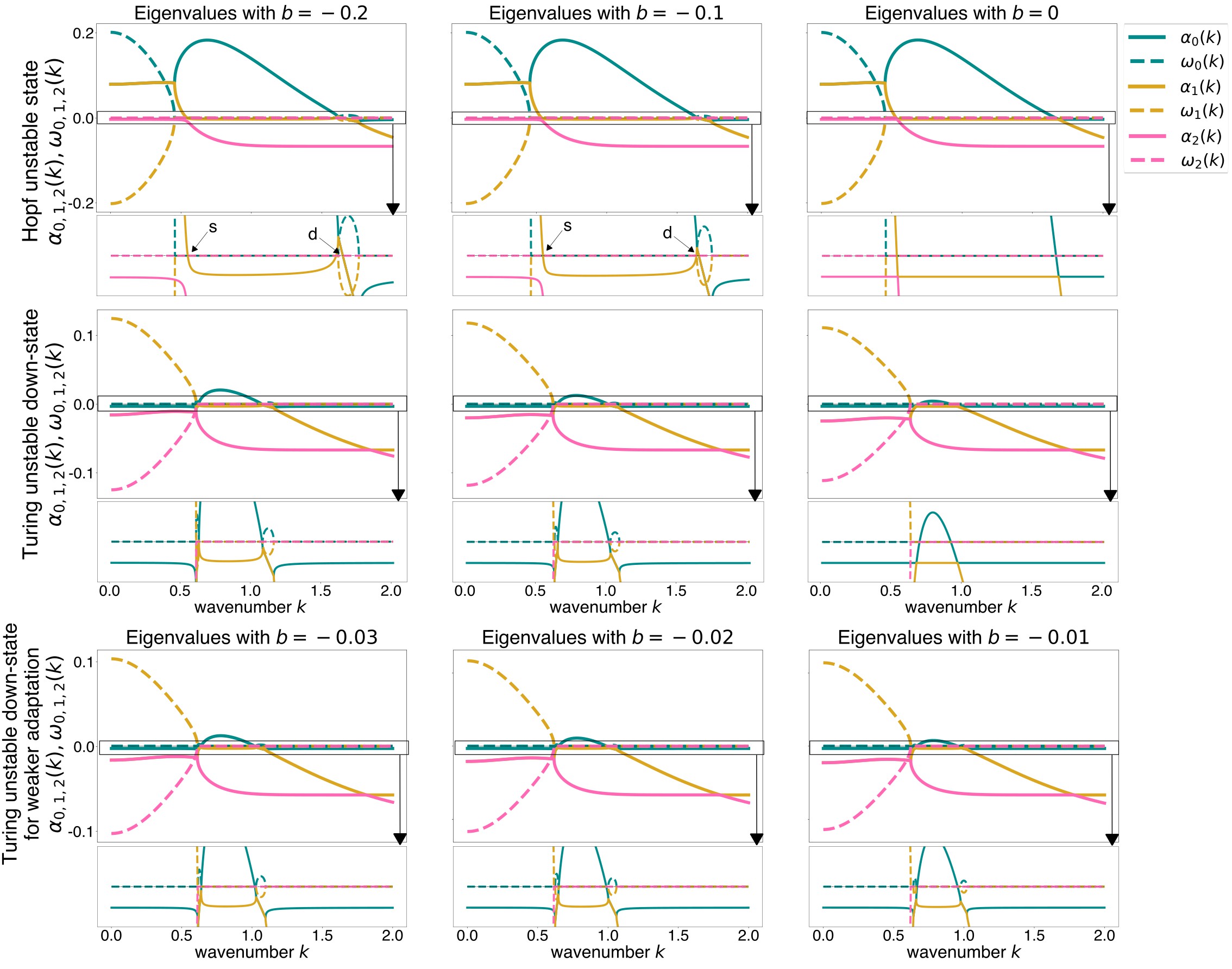}
    \caption{Eigenvalue spectrum for small values of $|b|$. The panels show  the real, $\alpha_{0,1,2}(k)$, (solid lines) and imaginary parts, $\omega_{0,1,2}(k)$, (dashed lines) of the eigenvalues of the linearization matrix in Eq.\ \eqref{eq:spatial-adaps-jacobian}. Top row: Hopf-unstable state at $(I_e, I_i) = (-0.35, -1.225)$. Center row: Turing-unstable down-state at $(I_e, I_i) = (-0.35, -0.55)$. Bottom row: same Turing-unstable down-state as in the center row but with weaker adaptation. Adaptation strength $|b|$ decreases from left to right. Lower panels show close-ups around the zero crossing of the real parts. The imaginary parts vanish for $b=0$. Arrows in panels below the first two plots of the upper row indicate zero crossing of a static (\textit{s}) and dynamic (\textit{d}) Turing instability.}
    \label{fig:eigenvalues}
\end{figure*}

Figure \ref{fig:equivalency-results-with-traces} shows a comparison between the state spaces and activity traces for the systems with SFA and h-currents. We see in Fig.\ \crefformat{figure}{#2#1{(a)}#3}\cref{fig:equivalency-results-with-traces} that the dynamical regimes are equivalent for SFA (left) and h-currents (right), except for the shift to the right (left) by the absolute value $|b|$ for SFA (h-currents), consistent with the required compensation of external inputs to the excitatory population (see Section \ref{sec:equivalency-transformation}). The rightward shift of the dynamical regimes for SFA compared to h-currents implies that the system requires more external input to the excitatory population to produce activity patterns beyond the spatially and temporally homogeneous steady states. We also see that equivalent positions (considering the required compensatory inputs) cause activity patterns to be equivalent under the transformation $T$, i.e.,\ for $m\leftarrow 1-m$, as exemplified in Fig.\ \crefformat{figure}{#2#1{(b)}#3}\cref{fig:equivalency-results-with-traces}. 

For the investigated parameterization (see Table \ref{tab:default-params}), adaptation is required to cause the purely temporal or spatial patterns observed at $b=0$ to travel. We exemplify this in Fig.\ \ref{fig:eigenvalues} where we show the real ($\alpha_j(k)$) and imaginary ($\omega_j(k)$) parts of all three eigenvalues $\lambda_j(k)=\alpha_j(k)\pm i\omega_j(k)$ of the Jacobian (Eq.\ \eqref{eq:spatial-adaps-jacobian}) for a Hopf-unstable (upper row) and a Turing-unstable (center and bottom row) state. In the upper two rows, for small values of $b$, the zero crossings of the real parts of any eigenvalue coincide with non-zero imaginary parts, resulting in a dynamic Turing instability and enabling spatio-temporal patterns in that very state. For $b=0$ the imaginary part vanishes, resulting in a static Turing instability. However, adaptation can not be arbitrarily weak and must be stronger than some threshold to show a dynamic Turing instability. In the bottom row of Fig.\ \ref{fig:eigenvalues}, for example, the Turing instability is static for $b=-0.02$ and $b=-0.01$ allowing the system to produce standing waves (see Appendix \ref{appendix:activity-traces-for-finite-b}, Fig.\ \ref{fig:activity-traces-for-finite-b}). While the Turing instability at $k_0$ becomes dynamic for $b=-0.03$, all eigenvalues have negative real parts for $b=-0.02$ for every $k$ for which the imaginary part is non-zero.

\begin{figure*}[t]
    \centering
    \includegraphics[width=0.99\linewidth]{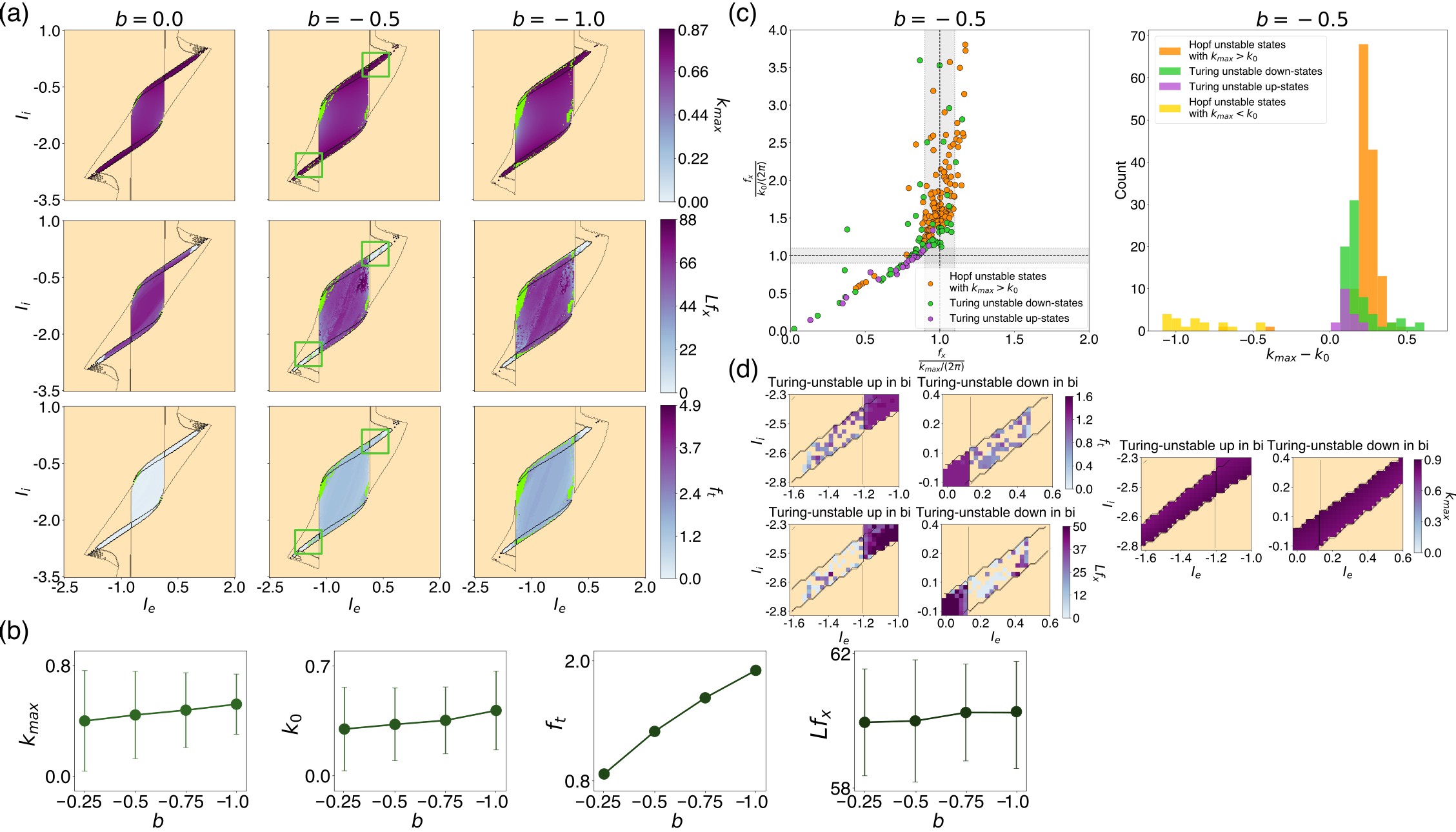}
    \caption{Properties of the spatio-temporal patterns emerging in Hopf- and Turing-unstable states for increasing h-current strength. \textbf{(a)} $k_{max}$ values (top row), number $L \cdot f_x$ of bumps on the ring (middle row), and dominant temporal frequency $f_t$ (bottom row) for each location in state space. The figure shows slices of state space spanned by the external input $(I_e, I_i)\in[-2.5,2]\times[-3.5,1]$ for $b\in\{0,-0.5, -1.0\}$. Purple colors denote the corresponding feature values. Darker colors indicate higher values, see color bar. Yellow denotes states of no spatio-temporal activity, light green denotes states with irregular patterns with a standard deviation $r$ of the Kuramoto-order parameter of $r>10^{-2}$. Thin lines denote the boundaries of the dynamical regimes shown in Fig.\ \ref{fig:equivalency-results-with-activity}\rm{(a)}. Green boxes correspond to the panels shown in \rm{(d)}. \textbf{(b)} Mean (dots) and variance (bars) of $k_{max},\ k_{0},\ f_t,$ and $L\cdot f_x$ for $b\in\{-0.25, -0.5, -0.75, -1\}$ for the states located in the Hopf-unstable, Turing-unstable up-, and Turing-unstable down-state regimes, for regular spatio-temporal activity for which the standard deviation of the Kuramoto-order parameter over time and space was $r\leq 10^{-2}$ (see Appendix \ref{subsec:regularity-methods}). \textbf{(c)} Left: Scatter plot of the ratios between the spatial frequency $f_x$ of the emerging patterns and the spatial frequencies corresponding to the wavenumbers $k_0$ and $k_{max}$ for all locations in state space from (a) with $b=-0.5$, which were identified as Hopf-unstable states with $k_{max}\geq k_0$ (orange), Turing-unstable down- (purple), or Turing-unstable up-states (green). Perfect agreement with a given mode corresponds to a ratio of $1$ (vertical and horizontal dashed lines). The gray regions around each ratio being $1$ denote deviations smaller than $10\%$. Right: Distribution of the differences $k_{max}-k_0$ for the same state space, additionally including the Hopf-unstable states with $k_{max} < k_0$. \textbf{(d)} Close-ups of state space (see green boxes in \rm{(a)}) with values for $k_{max}$ (right), $L\cdot f_x$ (lower left), and $f_t$ (upper left) in the Turing-unstable up- (left panels) and down-states (right panels) in bi. Yellow denotes states of no spatio-temporal activity or with irregular patterns with a standard deviation $r$ of the Kuramoto-order parameter of $r>10^{-2}$. The other colors denote feature values (see color bars). Wavenumbers were acquired as described in Section \ref{sec:3d-stability-analysis-methods}; spatial and temporal frequencies as described in the Appendices \ref{subsec:initialisation-methods} and \ref{subsec:frequency-methods}. Numerical simulations were conducted with initialization close to the corresponding unstable fixed points.}\label{fig:simulation-results}
\end{figure*}

Additionally, we see in Fig.\ \ref{fig:eigenvalues}, as $|b|$ increases, the imaginary part becomes larger which indicates faster temporal oscillations, since each mode is given by 
\begin{align*}
    \exp(\lambda(k)t)\ \exp(ikx)=&\ \exp((\underbrace{\alpha(k)}_{=0}\pm i \omega(k))\ \exp(ikx)\\
    =&\ \exp(\pm i\underbrace{\omega(k)}_{\mathclap{\omega\uparrow\text{ for }|b|\uparrow}}t)\ \exp(ikx).
\end{align*}Above conclusions are confirmed for the entire slices of state space in Appendix \ref{appendix:static-dynamic-distinction}, Fig.\ \ref{fig:distinct bifurcations}.
\section{Characterization of regular spatio-temporal patterns in Hopf- and Turing-unstable states}\label{sec:simulation-results}
Figure \ref{fig:simulation-results} shows the results of the numerical investigation for the system with h-currents with increasing adaptation strength for the regimes of Hopf or Turing instability, where regular patterns occur. We did not consider the regimes of three fixed points (corresponding to dark and light beige in Fig.\ \crefformat{figure}{#2#1{(a)}#3}\cref{fig:equivalency-results-with-activity}), because numerical simulations close to the fixed points undergoing a Turing instability lead to irregular propagation of activity fronts similar to the pattern exemplified in Fig.\ \crefformat{figure}{#2#1{(b)}#3}\cref{fig:equivalency-results-with-activity}, panel 7, which shows irregular activity emerging in a Turing-unstable up-state in bi.

Figure \crefformat{figure}{#2#1{(a)}#3}\cref{fig:simulation-results} shows the wavenumber $k_{max}$ (top row), the spatial frequency multiplied to the length of the ring, $L\cdot f_x$ (center row, effectively describing the number of bumps on the ring) and temporal frequency, $f_t$ (bottom row) of the emerging patterns for different locations in state space and for increasing adaptation strength. $k_{max}$ values vary continuously across state space. Values are larger than $0.44$ with highest values occurring in Turing-unstable states (see color bar). The spatial frequency of the wave patterns change smoothly across state space for $b=0$ with the highest values occurring toward the center of the Hopf regime. Values decrease toward the borders. For $b=-0.5$ and $b=-1$, the distribution of $L\cdot f_x$ becomes less smooth with non-equidistant, irregular patterns ($r>10^{-2}$) emerging toward the borders of the Hopf regime, where the regimes of Turing instability, one stable fixed point, Hopf instability, and three fixed points with only one stable state meet (green areas inside instability regimes, Fig.\ \crefformat{figure}{#2#1{(a)}#3}\cref{fig:simulation-results}). The numerical simulations also confirm that the activity traces for $b=0$ are stationary spatial waves, as $f_t$ is zero. For adaptation strong enough, waves start moving resulting in a finite value of $f_t$ which increases with the value of $|b|$ (see also Fig.\ \crefformat{figure}{#2#1{(b)}#3}\cref{fig:simulation-results}) confirming what is expected from the mathematical analysis. For increasing $|b|$, the averages of the wavenumbers $k_{0}$, and $k_{max}$ increase in value, albeit the variance over state space is high. $f_x$ increases with $|b|$ consistent with the increase of the wavenumbers $k_0$ and $k_{max}$. Additionally, the average temporal frequency increases steeply, which is expected from the increasing imaginary parts at the zero crossings of the real parts of the eigenvalues as discussed in Section \ref{sec:state-spaces}. 

Figure \crefformat{figure}{#2#1{(c)}#3}\cref{fig:simulation-results}, right panel, shows the distribution of the differences $k_{max}-k_0$. The case, $k_{max}-k_0 < 0$, occurs only in locations with fixed points identified as Hopf unstable and where irregular activity patterns emerge. Figure \crefformat{figure}{#2#1{(c)}#3}\cref{fig:simulation-results}, left panel, shows the ratio between the simulated spatial frequencies and the wavenumbers $k_0$ and $k_{max}$ plotted against each other. Perfect agreement between $f_x$ and a wavenumber $k$ would be indicated by the corresponding point falling onto the horizontal or vertical line of the ratios being $1$. All activity patterns emerging from locations in state space with $k_{max} < k_0$ are irregular and are not considered in the comparison to $f_x$. The gray regions around each ratio being $1$ denote deviations smaller than $10\%$. Since $k_{max}$ denotes the strongest growing unstable mode (compare Section \ref{subsec:wavenumber}), one would expect points to lie close to the vertical line in the center (for which $\frac{f_x}{k_{max} / (2\pi)}$ is close to one), which is indeed the case. However, some exceptions for Turing-unstable up- and down-states occur.

Since we observe bistability in Hopf-unstable states between spatially homogeneous oscillations and temporally stationary spatial oscillations (see Section \ref{sec:state-spaces}), we performed the same analysis as in Figs. \crefformat{figure}{#2#1{(a)}#3}\cref{fig:simulation-results} and \crefformat{figure}{#2#1{(c)}#3}\cref{fig:simulation-results} in Appendix \ref{subsec:random-init} but initialized close to zero. Results are similar and shown in Appendix \ref{subsec:random-init} Fig.\ \ref{fig:random-figure}. This figure also shows the $k_0$ values over slices of state space corresponding to Fig.\ \crefformat{figure}{#2#1{(a)}#3}\cref{fig:simulation-results}.

\begin{figure*}[t]
    \centering
    \includegraphics[width=0.99\linewidth]{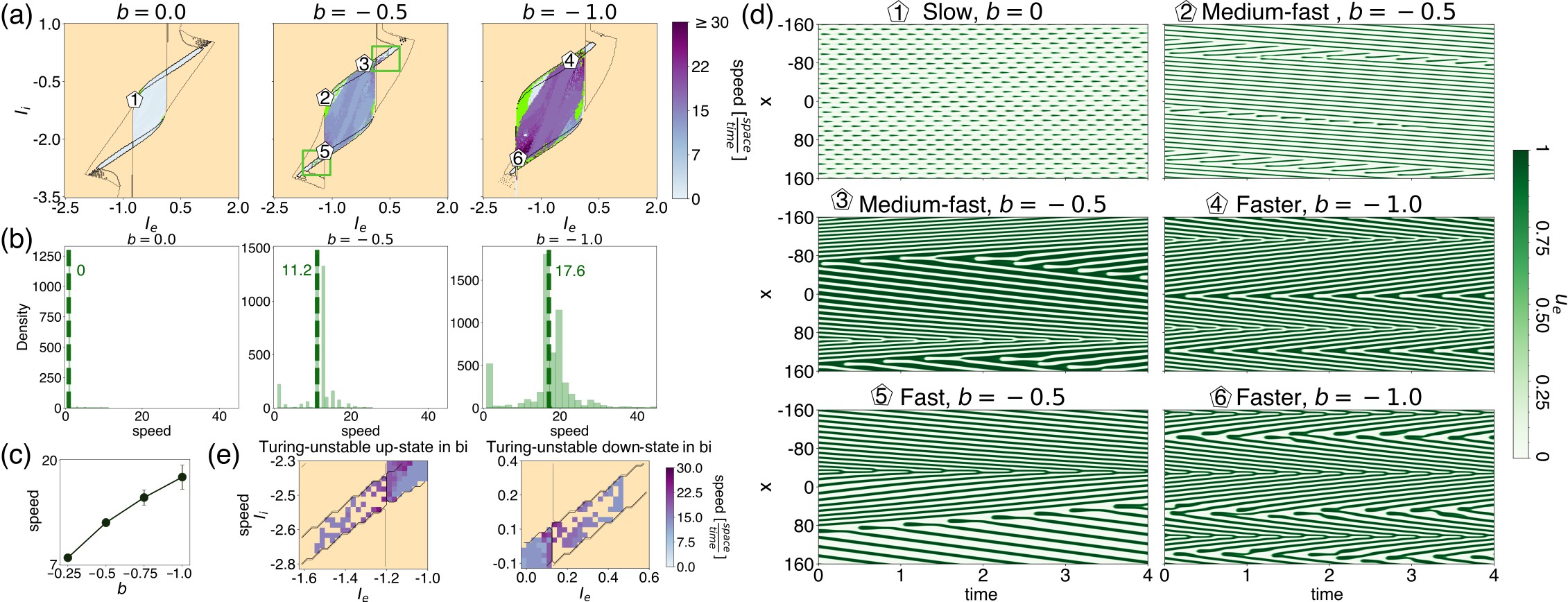}
    \caption{Propagation speed of spatio-temporal patterns emerging in Hopf- and Turing-unstable states for increasing adaptation strength. \textbf{(a)} Slices of state space spanned by the external input currents showing speed values for the corresponding location in state space. Adaptation strength increases from left to right. Speed values are indicated by purple (see color bar), yellow denotes states of no spatio-temporal activity, green denotes states of irregular patterns with $r>10^{-2}$. Thin lines denote the borders of the dynamical regimes shown in Fig.\ \ref{fig:equivalency-results-with-activity}\rm{(a)}. \textbf{(b)} Histograms of the distribution of speed values across the states shown in \rm{(a)}. Dashed vertical lines and numbers denote the location of the peaks of the histograms. \textbf{(c)} Average (dots) and variance (bars) of speed values across state space for $b\in\{-0.25, -0.5, -0.75, -1.0\}$. \textbf{(d)} spatio-temporal patterns of $u_e(x,t)$ for different values of $b$ emerging in the locations in state space which are indicated by hexagons in \rm{(a)}. The corresponding value of the external input currents are given in Table \ref{tab:speed-input-pairs} in Appendix \ref{appendix:state-space-activity-pairs}. The corresponding speed values are for 1: $<$0.5; 2 and 3: 11.6; 4: 17.7; 5: 14.6; 6: 18.7. All other parameters are given in Table \ref{tab:default-params}. \textbf{(e)} Close-ups of state spaces (see green boxes in Fig.\ \ref{fig:speed-results}\rm{(a)}) with values for speed. Numerical simulations were initialized around the fixed points of Hopf or Turing instability. Speed values were computed as described in Appendix \ref{subsec:velocity-methods}.}
    \label{fig:speed-results}
\end{figure*}

Figure \crefformat{figure}{#2#1{(d)}#3}\cref{fig:simulation-results} shows close-ups of Turing unstable up- and down-states in bi for $b=-0.5$. $k_{max}$ values do not deviate strongly from the $k_{max}$ values of the rest of the considered state spaces but only few numerical simulations converged to regular spatio-temporal patterns (non-yellow locations in the close-ups of state space). The resulting patterns strongly depend on the initialization, and even initializations close to the Turing-unstable fixed point do not necessarily converge to a regular spatio-temporal pattern but to the coexisting stable fixed point, as exemplified in Fig.\ \crefformat{figure}{#2#1{(b)}#3}\cref{fig:equivalency-results-with-activity}, panel 5(*), or irregular patterns\footnote{States marked in yellow in the regions of Turing-unstable up- and down-states in bi shown in Fig.\ \crefformat{figure}{#2#1{(d)}#3}\cref{fig:simulation-results} and Fig.\ \crefformat{figure}{#2#1{(e)}#3}\cref{fig:speed-results} may also support regular spatio-temporal patterns if another initialization would have been chosen.}, as exemplified in Fig.\ \crefformat{figure}{#2#1{(b)}#3}\cref{fig:equivalency-results-with-activity}, panel 7. If regular spatio-temporal patterns emerge, traveling waves show low values of spatial frequency. Temporal frequencies are also lower compared to the rest of state space.
To determine the propagation speed of adaptation-driven waves, we computed the phase velocity as described in Appendix \ref{subsec:velocity-methods} for each spatio-temporal activity pattern emerging in Turing- or Hopf-unstable states. Figure \crefformat{figure}{#2#1{(a)}#3}\cref{fig:speed-results} shows the absolute value of the velocity, i.e.,\ the \textit{speed} for different locations in state space and for $b\in\{0.0, -0.5, -1.0\}$. Speed values increase, the larger $|b|$ becomes. Toward the border of the regime of Hopf instability at which multiple different instability regimes meet (e.g.,\ see region around hexagons 1 and 2), the activity patterns are either not regular (i.e.,\ $r>10^{-2}$) or not spatio-temporal (e.g.,\ purely spatial). Toward the regions of Turing instability in bi, speed increases (e.g.,\ see darker colors around hexagon 6).

Figure \crefformat{figure}{#2#1{(b)}#3}\cref{fig:speed-results} shows the distribution of speed values across Hopf-unstable and Turing-unstable states. The stronger the adaptation, the higher the speed values are. For $b=0$, as expected, activity patterns do not travel across space (recall Fig.\ \ref{fig:equivalency-results-with-activity}, panel 0) or their speed is $<0.5$ (see Fig.\ \crefformat{figure}{#2#1{(d)}#3}\cref{fig:speed-results}, panel 1). For $b=-0.5$, there are two different speed values appearing the most often, with the dominant speed being $11.2$. For $b=-1$, the variance of the distribution is larger and the most common speed value is once again higher at $17.6$. Figure \crefformat{figure}{#2#1{(c)}#3}\cref{fig:speed-results} confirms that the average speed across state space increases with adaptation strength. Figure \crefformat{figure}{#2#1{(d)}#3}\cref{fig:speed-results} shows traveling waves for the same relative locations in state space (i.e.,\ considering the shift to the left discussed in Section \ref{sec:state-spaces}), increasing in adaptation strength from left to right column. ``Broad'' patterns become ``narrower'', increasing both temporal and spatial frequencies and their slope in the space-time diagrams, henceforth, the speed.

Figure \crefformat{figure}{#2#1{(e)}#3}\cref{fig:speed-results} shows the same close-ups as Fig.\ \crefformat{figure}{#2#1{(d)}#3}\cref{fig:simulation-results} of Turing-unstable up- and down-states in bi showing the speed values for the numerical simulations converging to regular traveling waves. Speed values tend to be higher closer to the boundary to the regimes of single Turing-unstable up- and down-states.
\section{Discussion}\label{sec:discussion}
We implemented the adaptive Wilson-Cowan field model in a homogeneous setting with local connectivity on a one-dimensional spatial domain and periodic boundary conditions. Each position on the ring consists of an excitatory and an inhibitory population of neurons. The excitatory population is equipped with an adaptation mechanism that either corresponds to a hyperpolarizing negative feedback current (SFA) or a hyperpolarization-activated positive feedback current (h-currents). The parametrization used throughout this study was chosen to display multiple different phenomena which are also observed in the brain - from excitation-inhibition-driven faster oscillations, regimes of bistability, to spatio-temporal activity patterns that show SO-like dynamics as either adaptation mechanism is active.

We investigated the model system for increasing adaptation strength $|b|$ only although the adaptation time constant $\tau_m$ has an effect on the temporal dynamics of adaptation-driven traveling waves as well. The larger $\tau_m$, the lower the dominant temporal frequency becomes (results not shown, but see \cite{Augustin2013}) as has also been observed experimentally (see \cite{Giocomo2008}). Changing the time constant scales the eigenvalues for the different parameterizations explored. This has no effect on the shape and size of the (in-)stability regions, but can influence the properties of the zero-crossing (i.e.,\ whether the root at $k_0$ is real or whether it has a non-zero imaginary part) and the type of the emerging activity patterns (i.e.,~ stationary vs.\ spatio-temporal, results not shown). This is consistent with \cite{Curtu2004} who showed that for a reduced system, $\tau_m$ determines whether a Turing bifurcation is dynamic or static. \vskip-1mm

A semi-analytical stability analysis across parameter space was performed for increasing adaptation strength, resulting in the identification of over ten dynamical regimes. These included regimes of bistability between two stable fixed points and a regime that admits bi- (or potentially multi-) stability between different activity patterns depending on the initialization. The dynamical landscape exhibits parallels with the models studied in \cite{Torao2021, Dimulescu2025, Levenstein2019, Bressloff2010}, but our analysis additionally displays regimes with Turing-unstable states, which are located at the border of regimes exhibiting oscillatory activity or a bistability between two fixed points. Spatio-temporal patterns, most often periodic adaptation-driven traveling waves, emerge in states exhibiting Hopf and Turing instabilities if adaptation is sufficiently strong. Activity patterns remain stationary in the absence of adaptation which is explained by the zero imaginary part of all eigenvalues of the linearization matrix when a fixed point displays a Turing instability. This aligns with the results presented in \cite{Curtu2004}, who proved the existence of stationary patterns for the reduced system when adaptation was sufficiently weak, while traveling waves occurred for stronger adaptation. 

Changing the adaptation strength not only alters the model's dynamical regimes but also the features of the emerging temporal and spatial patterns including the speed of traveling wave activity. The numerical simulations aligned with the mathematical analysis performed before. With stronger adaptation, wavenumbers increase on average, causing more waves on the ring, i.e.,\ higher dominant spatial frequencies. Analogously, with growing imaginary parts of the eigenvalues at the zero crossing of their respective real parts for stronger adaptation, temporal frequencies also increase. These changes correlate with increasing traveling speeds of activity patterns when the adaptation becomes stronger. In our two-population model, adaptation is a modulatory factor for spatio-temporal patterns, contrary to the purely excitatory model with synaptic depression of \cite{Kilpatrick2010}.

Our results also show that the two adaptation mechanisms are mathematically equivalent when adaptation strength is compensated by a changed external input current. This is not a contradiction to the results of \cite{Mehrotra2024} who investigated SFA versus h-currents experimentally and using computational modeling. They identified differences between SFA and h-currents induced slow sequences of synchronized neuronal activity and sequential state transitions between up- and down-states. While a spatial gradient in the adaptation strength for SFA caused sequential down-state onsets and synchronized up-state onsets, h-currents replicated experimentally observed synchronized down-state onsets and sequential up-state initiation. Here, we show that the adaptation strength $b$ has to be compensated by an additional external input to the excitatory population to preserve equivalency all the while adaptation induces and governs the emergent spatio-temporal properties. Consequently, a spatially heterogeneous adaptation strength (i.e.,\ $b\rightsquigarrow b(x)$) will shape the emergent spatio-temporal activity patterns over space and break the equivalence between SFA and h-currents unless it is compensated by a spatially dependent external input $I_e \rightsquigarrow I_e(x)$.

In our population-based model, both adaptation mechanisms were implemented in a simplified way. Hyperpolarizing feedback current of somatic SFA can be mediated by potassium channels, which activate when the excitatory neurons are in a depolarized high-activity state, effectively causing K$^+$ outward flow \cite{McCormick1992, Bhattacharjee2005}. Different potassium channels have different properties leading to differential effects in the actual adaptation mechanism: Voltage- ($I_m$, see \cite{BrownAdams1980}) and sodium-activated K$^+$-channels ($I_{KNa}$, see \cite{Schwindt1989}) contribute mostly to subtreshold, while Ca$^{2+}$-activated K$^+$-channels ($I_{KCa}$, \cite{BrownGriffith1983}) contribute to spike-based adaptation. Different channel types also have different effects on the response properties of a neuron, for example, on a neurons f-I curve when adaptation strength is increased \cite{Ladenbauer2014}. In addition, SFA can be mediated by sodium current inactivation which cause the excited population of neurons to become less depolarized via the inactivation of Na$^+$ inward flow \cite{Upchurch2022}. That said, our SFA model appropriately captures their canonical role as slow, hyperpolarizing feedback mechanisms which activate in an up-state. H-current based adaptation is mediated by a different channel dynamics, causing dichotomous effects. The responsible hyperpolarization-activated cyclic nucleotide gated (HCN) channels cause K$^+$ outward flow, all the while not preventing Na$^+$ inward flow \cite{Lee2017}, and have a reversal potential between -50 and -20 mV \cite{Combe2021Review}. Unlike potassium channels who have a clear selection of K$^+$ (see \cite{Mironenko2021}), HCN channels have a permeability of 3:1 to 5:1 for K$^+$:Na$^+$ \cite{Combe2021Review, Pape1996, Biel2009}. HCN channels are active as long as the neurons conductance is close to its resting membrane potential, causing positive inward currents that effectively excite the neuron. This, we simulate here, whereas the stagnating inhibitive effect for higher values of the membrane potential is neglegted similar to the cortical population model of \cite{Mehrotra2024}. Clearly, dynamical equivalence between SFA and h-current based adaptation is only expected to strictly hold for the approximate description of the adaptation dynamics given by Eqs.\ \eqref{eq:model} and \eqref{eq:tranfer-function}.

For the reason of particularly addressing the influence of the dichotomy of the most prominent features of the adaptation mechanisms, we omitted delays, although it has been shown that delays can drive directions of waves in oscillator networks (see \cite{Budzinski2023}) and enrich the emergent dynamics of complex spatio-temporal patterns (see \cite{Roberts2019}). Delays however would not affect the shown symmetry of the adaptation mechanisms.

Traveling waves of SOs are often presented as the propagation of alternating up- and down-states (see \cite{Torao2021, Sanchez-Vives2020}), in which one can separate the propagation of up-states via down-to-up transitions from the traveling of down-states via up-to-down transitions. In our model, we show that h-currents require less external input to the excitatory population than SFA to transition from a maintained stable down-state into oscillatory states, due to the shift of state space to the left along the $I_e$-axis. Note that even though we conduct simulations in a parameter space spanning negative values for external input currents, including thresholds in the transfer functions of the neuronal populations shifts the state space, exactly as is, along the axes of the external input currents by the value of the thresholds. 

In the mean-field model with SFA of \cite{Cakan2020}, the authors show that SFA shifts the state space horizontally to the right along the axis of the external input to the excitatory population, which agrees with our results. Consequently, to simulate a maintained up-state activity that, either by transient or noisy inputs, allows temporal deactivation of the neuronal populations into a down-state (i.e.,\ down-state onsets), requires more external input to the excitatory population. Additionally, we see in our state spaces that the opposite holds for h-currents, induced by the shift to the left. H-currents consequently require less external input both for up-state onsets as well as down-state onsets, and therefore, also lower transient inputs to maintain a slow alternation between up- and down-states.
\vspace{-4mm}

Further studies could extend our investigations to a biophysically more realistic model, for example, as in \cite{Ladenbauer2014}, which distinguishes subthreshold activated versus spike-triggered SFA. \cite{Augustin2013} show for the model of \cite{Ladenbauer2014} that subthreshold activated SFA changes the neuronal input-output relationship, and can increase the variability of interspike intervals (ISI) independent of current inputs, whereas spike-triggered SFA changes the ISIs variability based on the input to the neurons. It would be interesting to see, how far the diverse types of activations in SFA would break the shown symmetry to h-current based adaptation. For both adaptation mechanisms, the work of \cite{Porta2024, MartinPedersen2024} for non-spatial conductance based-models would allow to simulate other potentially symmetry-breaking channel dynamics, e.g.,\ the stagnating inhibitive effect of h-currents. The Wilson-Cowan model investigated in this study may serve as a reference when exploring the effects of a biophysically realistic channel dynamics.


\begin{acknowledgments}
We thank Prof.\ Tilo Schwalger (Technische Universit\"{a}t Berlin) for very helpful comments and discussions. This work was funded by the Deutsche Forschungsgemeinschaft (DFG) as part of the CRC 1315 (project number 327654276). It was published on arXiv as a preprint in Oct. 2025 \cite{Stroemsdoerfer-preprint}.
\end{acknowledgments}
\appendix

\section{}
\subsection{Finite bandwidth of unstable modes}\label{appendix:finite-bandwidthW}
The eigenvalues depend on the Fourier transform of the spatial kernel $w_j(x)\ j\in\{e,i\}$, Eq. \eqref{eq:kernel} which is a normalized Gaussian in the real domain. In the Fourier domain, it has its maximum magnitude at $0$ with $|\hat{w}_j(0)|=1$, and its amplitude $|\hat{w}_j(k)|$ decays to zero for $k\to\infty$. Furthermore, the dynamical system is well-defined, since the sigmoidal transfer functions Eq. \eqref{eq:tranfer-function} restrict the state variables always to the range between $[0,1]$, and their derivatives are bounded functions. As long as the time constants $\tau_j,\ j\in\{e,i,m\}$, are non-zero, the eigenvalues remain finite for any wavenumber $k$. Infinitely sized bands of unstable modes can not exist because the Jacobian matrix \begin{widetext}\begin{align*}
    A_3(k)=\begin{bmatrix}
        -\frac{1}{\tau_e}+\frac{w_{ee}}{\tau_e}F^\prime_e(B_e)\hat{w}_e(k) & -\frac{w_{ei}}{\tau_e}F^\prime_e(B_e)\hat{w}_i(k)& -\frac{b}{\tau_e}F^\prime_e(B_e) \\
        \frac{w_{ie}}{\tau_i}F^\prime_i(B_i)\hat{w}_e(k) & -\frac{1}{\tau_i}-\frac{w_{ii}}{\tau_i}F^\prime_i(B_i)\hat{w}_i(k) & 0 \\
        \frac{F^\prime_m(\tilde{u}_e-\mu)}{\tau_m} & 0 & -\frac{1}{\tau_m}
    \end{bmatrix} \to \begin{bmatrix}
        -\frac{1}{\tau_e} & 0 & -\frac{b}{\tau_e}F^\prime_e(B_e) \\
        0 & -\frac{1}{\tau_i} & 0 \\
        \frac{F^\prime_m(\tilde{u}_e-\mu)}{\tau_m} & 0 & -\frac{1}{\tau_m}
    \end{bmatrix},\ \text{for }k\to\infty,
\end{align*}\end{widetext}has always eigenvalues with negative real parts in the long wavelength limit. The coefficients of the characteristic polynomial become \begin{align*}
    c_0 &= \frac{1 + bF_m^\prime(\tilde{u}_e-\mu)F_e^\prime(B_e)}{\tau_e\tau_i\tau_m} \\
    c_1 &= (\frac{1}{\tau_e\tau_i} + \frac{1}{\tau_i\tau_m} + \frac{1 + bF_m^\prime(\tilde{u}_e-\mu)F_e^\prime(B_e)}{\tau_e\tau_m})\\
    c_2 &= \frac{1}{\tau_e} + \frac{1}{\tau_i} + \frac{1}{\tau_m},
\end{align*}where $(\tilde{u}_e, \tilde{u}_i, \tilde{m})$ denotes a fixed point and $B_e$ is defined as for Eq. \eqref{eq:spatial-adaps-jacobian}. Since $b\cdot F_m^\prime(\tilde{u}_e-\mu)>0$ for both the SFA and the h-current settings, we have $c_0, c_1, c_2 > 0$. Furthermore, \begin{widetext}\begin{align*}
    c_2c_1 - c_0 = \underbrace{(\frac{1}{\tau_e} + \frac{1}{\tau_m})}_{> 0}(\underbrace{\frac{1}{\tau_e\tau_i} + \frac{1}{\tau_i\tau_m}}_{> 0} + \underbrace{\frac{1+bF_m^\prime(\tilde{u}_e-\mu)F_e^\prime(B_e)}{\tau_e\tau_m}}_{> 0}) + \underbrace{\frac{1}{\tau_e\tau_i^2} + \frac{1}{\tau_i^2\tau_m}}_{> 0}>0.
\end{align*}\end{widetext}As a consequence of the Routh-Hurtwitz stability criterion, all eigenvalues of the Jacobian matrix in the small wavelength limit have negative real parts. 
Therefore, all bands of unstable modes are of finite bandwidth.
\vspace{-4mm}
\subsection{Slices of state spaces for small values of $b$}\label{appendix:static-dynamic-distinction}
\begin{figure}[h!]
    \centering
    \includegraphics[width=0.90\linewidth]{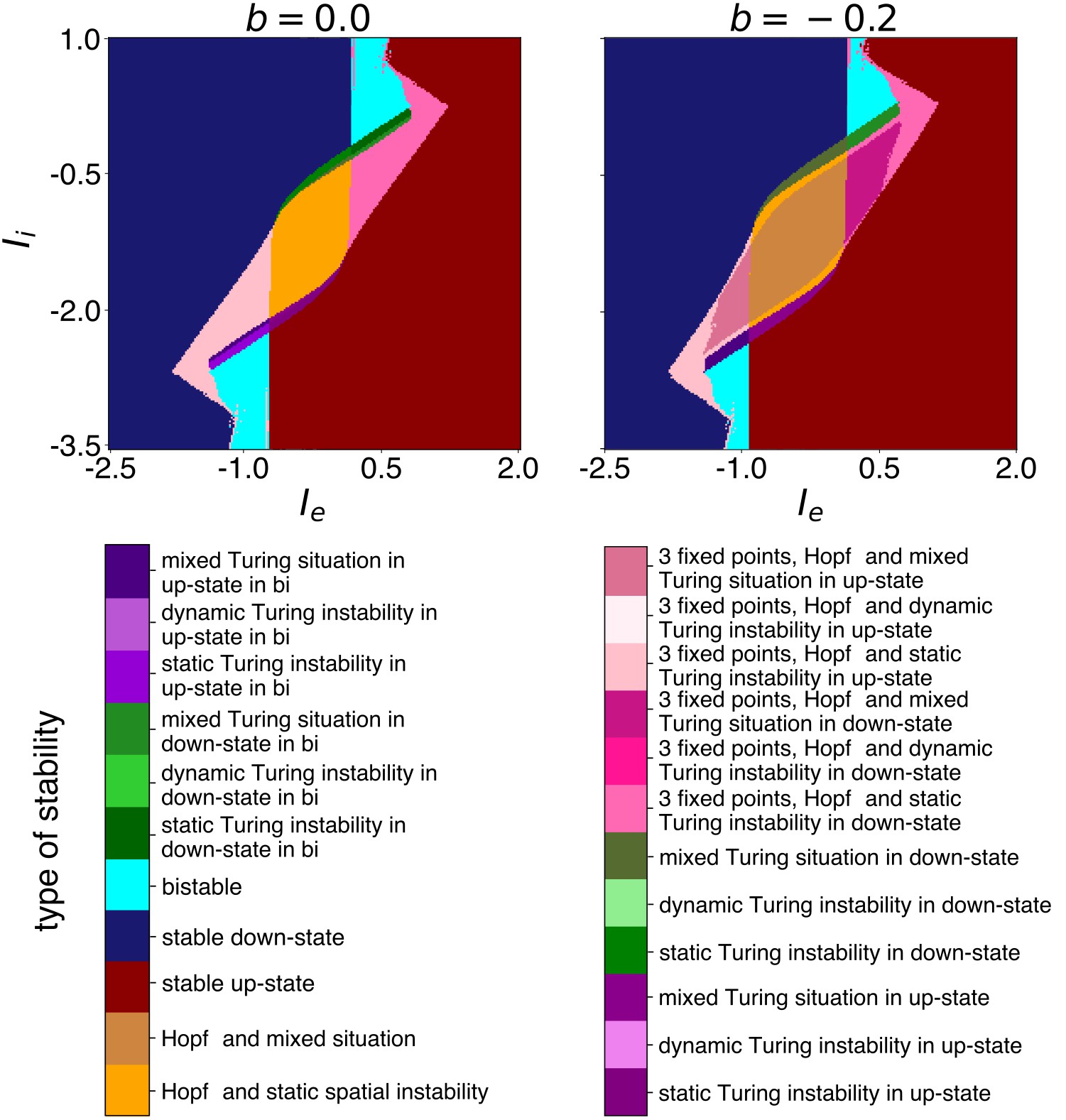}
    \caption{State spaces over input pairs $(I_e, I_i)\in[-2.5,2]\times[-3.5,1]$ for $b\in\{0,-0.2\}$. Colors denote the different types of instability and the corresponding fixed point in which they occur (see color bar).}
    \label{fig:distinct bifurcations}
\end{figure}
Figure \ref{fig:distinct bifurcations} shows the state spaces for $b=0$ and $b=-0.2$ including the distinction of static vs.\ dynamic Turing instability. Without adaptation, the system shows static Turing instabilities only. For weak adaptation ($b=-0.2$), nearly all states formerly exhibiting a static Turing instability display either a dynamic Turing instability or a mixed Turing situation. 
\subsection{Activity traces for traveling and standing waves with finite adaptation}\label{appendix:activity-traces-for-finite-b}
\begin{figure}[h!]
    \centering
    \includegraphics[width=0.99\linewidth]{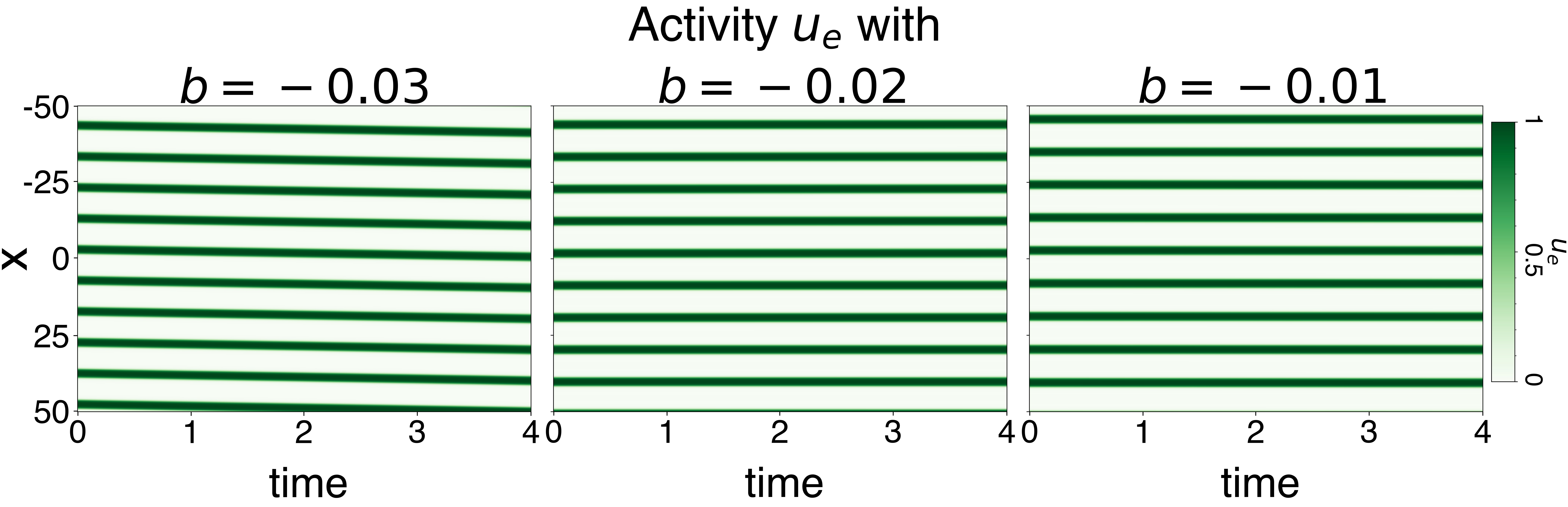}
    \caption{Activity traces $u_e(x,t)$ corresponding to the Turing-unstable down-state shown in Fig.\ \ref{fig:eigenvalues}, bottom row, for $x\in[-50,50]$.}
    \label{fig:activity-traces-for-finite-b}
\end{figure}
Figure \ref{fig:activity-traces-for-finite-b} shows the excitatory activity for very weak adaptation. While for $b=-0.03$ the waves travel over space, they become stationary for $b=-0.02$

\section{Numerical Simulations and Feature Acquisition}\label{appendix:numerical-simulations}
\subsection{Initialization and implementation}\label{subsec:initialisation-methods}
All models are implemented in \texttt{python}. The code is provided at \url{https://github.com/ronja-roevardotter/equivalent-adaptation-wilson-cowan-field.git}\ \footnote{Within this framework, notebooks generating Figs. \ref{fig:wavenumbers} and \ref{fig:eigenvalues} are provided, allowing users to zoom in for a more detailed view.} \cite{StroemsdoerferFramework}. The temporal integration is implemented using Runge-Kutta of the fourth order with an integration time step $dt=0.1$. For the spatial convolution, we used the Fourier transform of \texttt{numpy}. For every parameterization, we conducted numerical simulations with two different initializations in fixed points that were either Hopf- or Turing-unstable. For random initialization, we sampled $u_e(x,0), u_i(x,0), m(x,0)$ for every position $x$ on the ring from a uniform distribution $\mathcal{U}(0,0.001)$. When initializing close to the fixed point value $(\tilde{u}_e, \tilde{u}_i, \tilde{m})$, additive noise was sampled from a normal distribution with standard deviation 0.1. For the latter, we conducted three simulations with three different seeds, to ensure that all results presented in this study were not an artefact of the particular seed used. 
All features (temporal and spatial frequencies, regularity, and speed of spatio-temporal patterns) that were investigated were computed for the excitatory activity traces $u_e$. We conducted numerical simulations with a duration of $30$ time units and discarded the first $10$ time units to reduce the effects of transients. 

\subsection{Robustness of numerical methods and model dynamics}\label{appendix:robustness}
Fixed points were computed numerically at each location in state space using the Python function \texttt{scipy.optimize.root()} with more than 60 different initial conditions. All found fixed points $(\tilde{u}_e,\tilde{u}_i,\tilde{m})$ satisfied the root conditions (i.e., Eqs. \eqref{eq:matrix-model}, $S(\tilde{u}_e,\tilde{u}_i,\tilde{m}))=\vec{0}$) within an absolute tolerance of $<10^{ - 8}$. Stability was determined from analytically derived Jacobian matrices.

To assess the robustness of the stability analysis, eigenvalues were recomputed after perturbing each fixed point by $\xi=10^{-8}$, i.e., $(\tilde{u}_e+\xi, \tilde{u}_i+\xi, F_m(\tilde{u}_e+\xi-\mu))$ and $(\tilde{u}_e-\xi, \tilde{u}_i-\xi, F_m(\tilde{u}_e-\xi-\mu))$. The real part of the dominant eigenvalue, $\max_j(\real(\lambda_j(k))),\ j\in\{0,1,2\}$, determines the stability of a mode $k$. We now separately consider the modes $k=0$ and $k>0$. To identify whether the stability of a fixed point is correctly identified at $k=0$, or for some $k>0$, we computed \begin{align*}
    M_0(u_e, u_i, m) &= \max_j\bigl( \real(\lambda_j(0))\bigr) \\
    M(u_e, u_i, m) &= \max_{j,k}\bigl( \real(\lambda_j(k))\bigr)
\end{align*}for the fixed point and their perturbed values to then collect the minimum and maximum, i.e.,\ \begin{widetext}
    \begin{align*}
    \min(M_0) &= \min\{M_0(\tilde{u}_e,\tilde{u}_i,F_m(\tilde{u}_e-\mu)), M_0(\tilde{u}_e+\xi, \tilde{u}_i+\xi, F_m(\tilde{u}_e+\xi-\mu)), M_0(\tilde{u}_e-\xi, \tilde{u}_i-\xi, F_m(\tilde{u}_e-\xi-\mu))\}\\
    \max(M_0) &= \max\{M_0(\tilde{u}_e,\tilde{u}_i,F_m(\tilde{u}_e-\mu)), M_0(\tilde{u}_e+\xi, \tilde{u}_i+\xi, F_m(\tilde{u}_e+\xi-\mu)), M_0(\tilde{u}_e-\xi, \tilde{u}_i-\xi, F_m(\tilde{u}_e-\xi-\mu))\}\\
    \min(M) &= \min\{M(\tilde{u}_e,\tilde{u}_i,F_m(\tilde{u}_e-\mu)), M(\tilde{u}_e+\xi, \tilde{u}_i+\xi, F_m(\tilde{u}_e+\xi-\mu)), M(\tilde{u}_e-\xi, \tilde{u}_i-\xi, F_m(\tilde{u}_e-\xi-\mu))\}\\
    \max(M) &= \max\{M(\tilde{u}_e,\tilde{u}_i,F_m(\tilde{u}_e-\mu)), M(\tilde{u}_e+\xi, \tilde{u}_i+\xi, F_m(\tilde{u}_e+\xi-\mu)), M(\tilde{u}_e-\xi, \tilde{u}_i-\xi, F_m(\tilde{u}_e-\xi-\mu))\}. 
\end{align*}
\end{widetext}If $\min(M),\ \max(M)<0$, the fixed point is classified as stable. Vice versa, if $\min(M), \max(M)>0$, the fixed point is classified as unstable. Lastly, if $\min(M)<0$ and $\max(M)>0$, the classification would be ambiguous. This case most likely occurs in locations in state space which are close to a bifurcation. We found that for all locations inside the instability regions which we evaluated, $\min(M)>0$ was fulfilled with a minimum at $7.8\cdot10^{-6}>0$. We found the same for $\min(M_0)$ for all evaluated locations in state space inside the regime of Hopf instability. This implies that at all representative locations in state space for which we performed the stability analysis, the corresponding fixed points were robustly classified as unstable, and that, due to the chosen discretization of state space (181 $\times$ 181 grid), the evaluated parameterizations were sufficiently far from bifurcation lines. Since the perturbation value $\xi=10^{-8}$ corresponds to the minimum accuracy of the fixed points, these results give confidence that stability is robustly assessed.

Furthermore, we assessed the robustness of the type of the emerging activity patterns (i.e.,~whether a pattern is purely spatial, purely temporal, spatio-temporal or neither) when the system is initialized farther away from the unstable fixed points. We chose five different locations in state space to represent the range of dynamical regimes in this study. Let $\tilde{\vec{u}}=(\tilde{u}_e, \tilde{u}_i, \tilde{m})^\top$ be an unstable fixed point in either location in state space. We initialize
\begin{align*}
    \vec{u}_{init}(x) = \tilde{\vec{u}} \pm (\varepsilon,\varepsilon,\varepsilon)^\top + \vec{\xi}(x),
\end{align*}for every spatial position $x$ on the ring, where $\vec{\xi}(x)\in\mathbb{R}^3$ is independently drawn from a uniform distribution $\mathcal{U}(-0.01, 0.01)$ for each location. We chose the baseline value $\varepsilon\in\{0.01, 0.05, 0.1, 0.2, 0.3, 0.4\}$. We manually inspected each fixed point value $(\tilde{u}_e, \tilde{u}_i, \tilde{m})$ and adjusted the baseline accordingly: upwards (i.e., $+\varepsilon$) if $(\tilde{u}_e, \tilde{u}_i, \tilde{m}) \leq (0.42, 0.42, 0.42)$, and downwards (i.e., $-\varepsilon$) otherwise, so that $\tilde{\vec{u}} \pm (\varepsilon,\varepsilon,\varepsilon)^\top\in[0,1]^3$. Furthermore, we manually verified that the added noise $\vec{\xi}(x)$ was small enough that $\vec{u}_{init}(x)\in[0,1]^3$ was also true everywhere.

We conducted numerical simulations for five different realizations of $\vec{\xi}(x)$, consequently leading to $5\times6$ simulations per representative location in state space. This investigation showed that the variations in spatial frequencies and speed were comparably low, and that the temporal frequency remained at the same value for all simulations, independent of the initialization. For all states except the Hopf-unstable state, we observed that the resulting activity pattern was always spatio-temporal. For the Hopf-unstable state, however, a change of the baseline by $\varepsilon\geq0.05$ led to purely temporal oscillations (see also Chapter \ref{sec:simulation-results}).

\subsection{Spatial and Temporal Frequencies}\label{subsec:frequency-methods}
We define \textit{one time unit} as $unit_t = 10,000$ time-iteration steps and quantify temporal frequency by $\bigl[ \frac{oscillation\ periods}{unit_t} \bigr]$. For brevity, we write $\bigl[ \frac{osc}{unit_t} \bigr]$. This choice is based on the model's ability to generate slow oscillations (i.e.,\ $\leq 2 \frac{osc}{unit_t}$) for finite adaptation (i.e.,\ $b\neq0$) as well as faster oscillations (e.g.,\ $\alpha$-like oscillations with $8-13\ \frac{osc}{unit_t}$) that are generated by the excitatory-inhibitory loop (see, e.g.,\ $f_t$ in Fig.\ \crefformat{figure}{#2#1{(a)}#3}\cref{fig:random-figure}). 
To determine the dominant temporal frequency, we computed the power spectra $P_{sx}(f)$ of the time series of the excitatory activity for each position $x$. Then we averaged over all positions, \begin{align*}
    \bar{P}_{s}(f) = \frac{1}{N}\sum_{x=1}^NP_{sx}(f),
\end{align*} and determined the frequency of maximum power of the averaged power spectrum, i.e.,\  $f_t = \argmax_{f}\bar{P}_{s}(f)$. To characterize spatial oscillations, we computed the power spectrum over space $P_{xt}(f)$ for each time step $t\in[0,T\cdot unit_t]$, where $T=20$ denotes the amount of time units used after discarding the transients (see Section \ref{subsec:initialisation-methods}). Then, we averaged over all time steps, \begin{align*}
    \bar{P}_{x}(f) = \frac{1}{T\cdot unit_t}\sum_{t=1}^{T\cdot unit_t}P_{xt}(f),
\end{align*}and determined the spatial frequency of maximum power over the averaged power spectrum, i.e.,\ $f_x=\argmax_f\bar{P}_x(f)$. For both temporal and spatial frequencies, activity patterns exist for which the power spectrum shows multiple peaks. In those cases we report the temporal and spatial frequencies corresponding to the most prominent peak. 

In the text, we sometimes report the value $L \cdot f_x$ instead of $f_x$, which correspond to the number of bumps on the ring.

\subsection{Regularity of patterns}\label{subsec:regularity-methods}
We characterized the spatio-temporal pattern using the Kuramoto-order parameter. We computed the analytical signal representation of the activity $u_e(x,t)$ using the Hilbert-Transform\begin{align}
    z(x,t) = u_e(x,t)+i\mathcal{H}[u_e(x,t)],\label{eq:hilbert-trafo}
\end{align}where $\mathcal{H}[\cdot]$ denotes the Hilbert-transformation and $i$ the imaginary unit. Using the instantaneous phase\begin{align}
    \theta(x,t)=\rm{arctan}\left(\frac{Im(z(x,t)}{Re(z(x,t))}\right)\label{eq:inst-phase}
\end{align}we calculate the Kuramoto-order parameter $R(t)$ for each time step $t$ \begin{align*}
    R(t)=\left|\frac{1}{n}\sum_{x}e^{i\theta(x,t)}\right|
\end{align*}and define the regularity $r$ of a spatio-temporal pattern by its standard deviation\begin{align*}
    r = \sqrt{\frac{1}{T}\sum_{t=1}^T (R(t)-\bar{R})}\ ,\ \ \text{ }\ \bar{R}=\frac{1}{T}\sum_{t=1}^T R(t)
\end{align*}over a time interval $[0,T]$. We considered patterns to be regular if $r\leq 10^{-2}$. This threshold corresponds to the mean over the ten most irregular patterns found when compiling Fig.\ \crefformat{figure}{#2#1{(a)}#3}\cref{fig:simulation-results}.

\subsection{Phase velocity and speed}\label{subsec:velocity-methods}
\begin{figure}[t]
    \centering
    \includegraphics[width=0.9\linewidth]{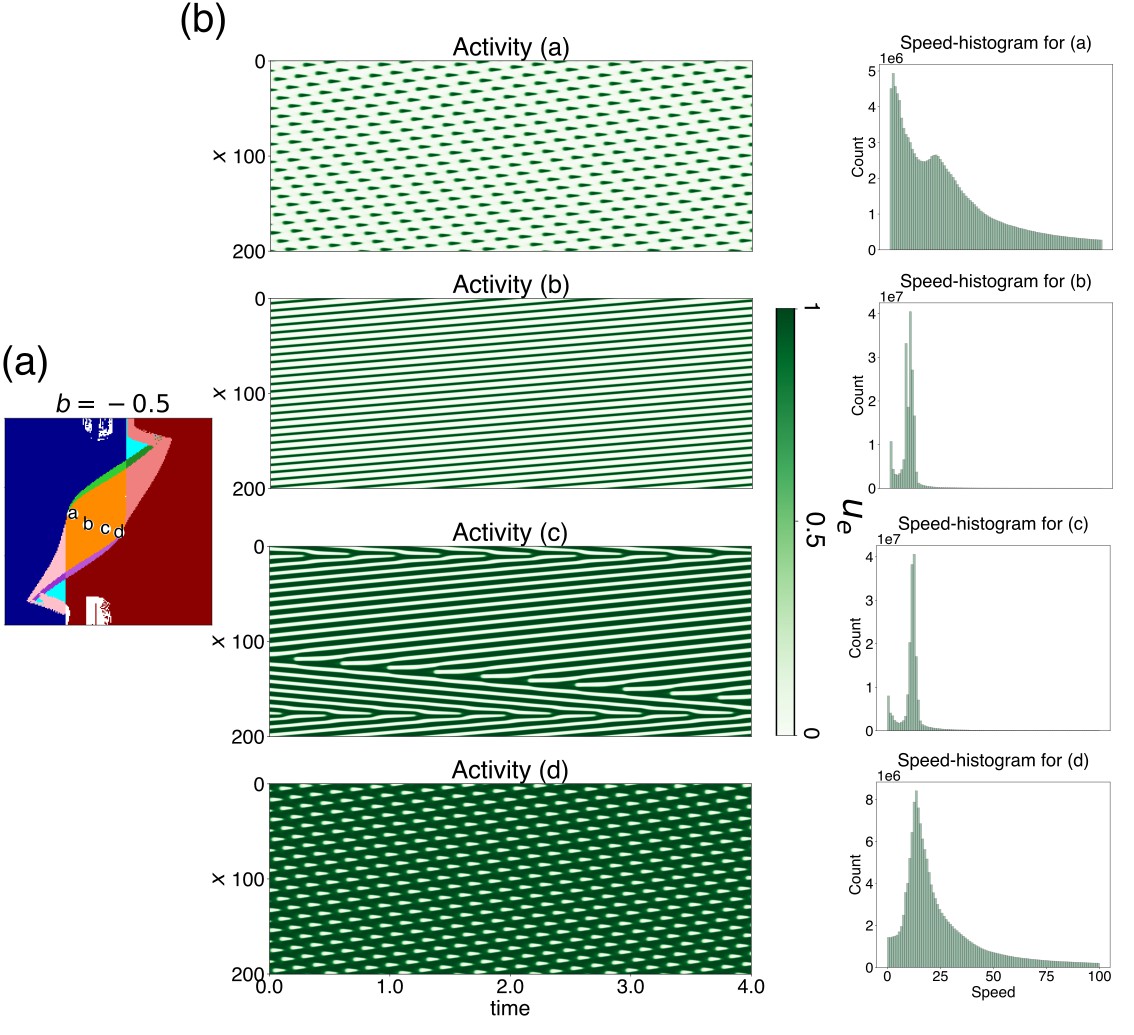}
    \caption{Computation of dominant speed values. \textbf{(a)} Locations in state space for the activity patterns (a)-(d) shown in panel \rm{(b)}. \textbf{(b)} Left panels: Activity traces for $x\in[0,200]$. Green colors denote activity values (darker green for higher, lighter green for low activity, see color bar). Right panels: Corresponding histograms of local speed values. The dominant speed corresponds to the speed value occurring most often.}
    \label{fig:speed-computation}
\end{figure}
To compute the speed in $\bigl[\frac{unit_x}{unit_t}\bigr]$, where $unit_x=\frac{1}{dx}$, of a spatio-temporal pattern, we used the approach of \cite{Rubino2006}. We first determine the analytical representation of the activity $u_e(x,t)$ as in Eq.\ \eqref{eq:hilbert-trafo}, where $\theta(x,t)$ is the instantaneous phase at position $x$ and time step $t$ as in Eq.\ \eqref{eq:inst-phase}. The local speed $s(x,t)$ is computed from the temporal and spatial derivatives $\frac{\partial\theta}{\partial x}(x,t)$ of the instantaneous phase,\begin{align*}
    s(x,t)=\left| \frac{\frac{\partial\theta(x,t)}{\partial t}}{\frac{\partial\theta(x,t)}{\partial x}} \right|.
\end{align*}Since spatio-temporal patterns may be composed of traveling waves of different speeds (e.g.,\ see Fig.\ \crefformat{figure}{#2#1{(b)}#3}\cref{fig:speed-results}, panel 5), we apply a similar approach as for the dominant spatial frequency. We compute a histogram over all local speed values appearing per activity pattern and take the speed value appearing the most often. In Fig.\ \ref{fig:speed-computation}, we illustrate this procedure for four different activity patterns.
\subsection{Properties of activity patterns emerging for initializations close to zero}\label{subsec:random-init}
\begin{figure}[h!]
    \centering
    \includegraphics[width=0.9\linewidth]{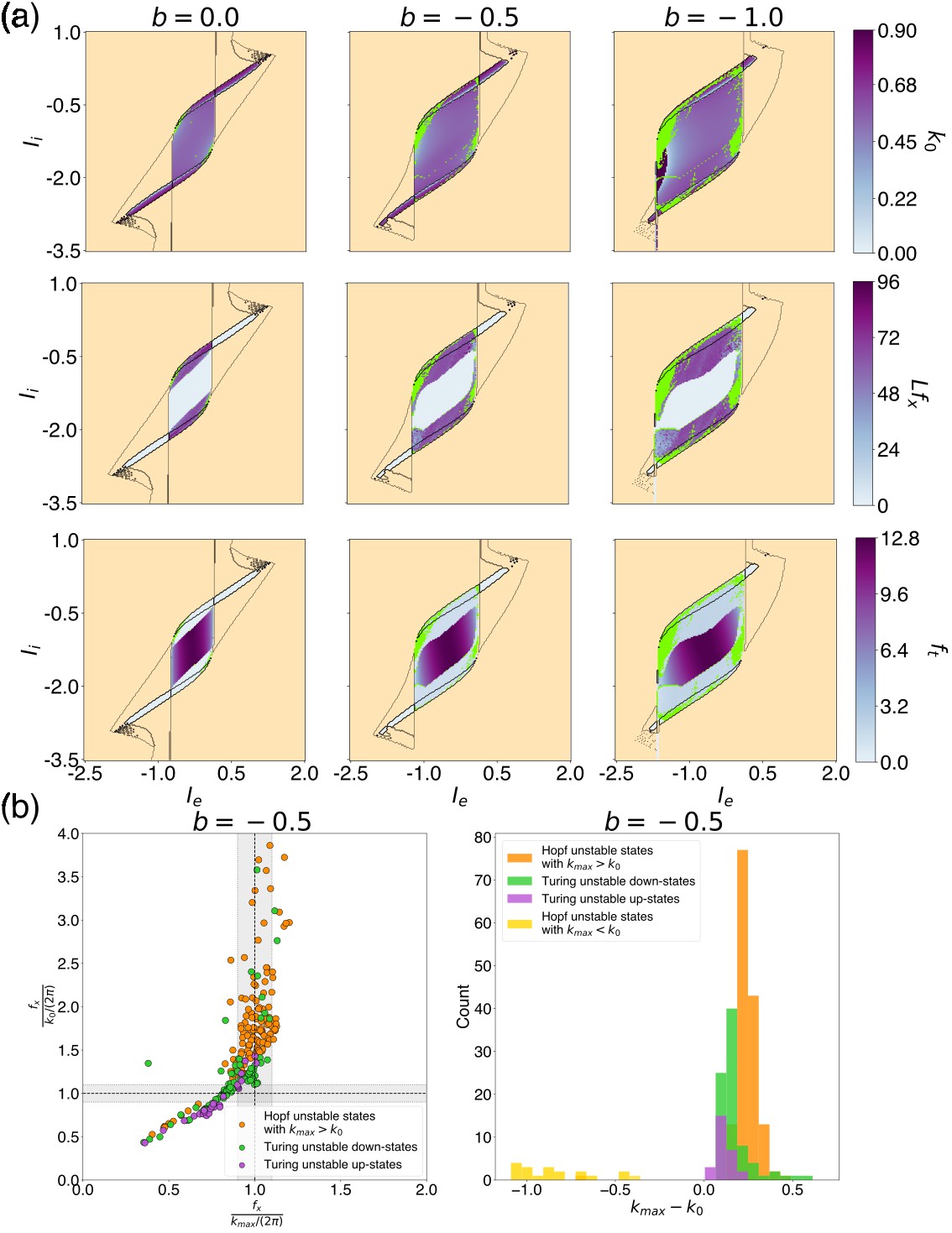}
    \caption{Properties of the spatio-temporal patterns emerging in Hopf- and Turing-unstable states for increasing h-current strength. \textbf{(a)} Wavenumber $k_{0}$ (top row), dominant spatial frequency $L\cdot f_x$ (middle row), and dominant temporal frequency $f_t$ (bottom row) for each location in state space. The figure shows slices of state space spanned by the external input $(I_e, I_i)\in[-2.5,2]\times[-3.5,1]$ for $b\in\{0,-0.5, -1.0\}$. Purple colors denote the corresponding feature values. Darker colors indicate higher values, see color bar. Yellow denotes states of no spatio-temporal activity, light green denotes states with irregular patterns with a standard deviation $r$ of the Kuramoto-order parameter of $r>10^{-2}$. Thin lines denote the boundaries of the dynamical regimes shown in Fig.\ \ref{fig:equivalency-results-with-activity}\rm{(a)}. \textbf{(b)} Left: Scatter plot of the ratios between the spatial frequency $f_x$ of the emerging patterns and the spatial frequencies corresponding to the wavenumbers $k_0$ and $k_{max}$ for all locations in state space from (a) with $b=-0.5$, which were identified as Hopf-unstable states with $k_{max}\geq k_0$ (orange), Turing-unstable down- (purple), or Turing-unstable up-states (green). Perfect agreement with a given mode corresponds to a ratio of $1$ (vertical and horizontal dashed lines). The gray regions around each ratio being $1$ denote deviations smaller than $10\%$. Right: Distribution of the differences $k_{max}-k_0$ for the same state space, additionally including the Hopf-unstable states with $k_{max} < k_0$. Numerical simulations were conducted with initializations close to zero.}
    \label{fig:random-figure}
\end{figure}

We conducted same analysis as for Figs. \crefformat{figure}{#2#1{(a)}#3}\cref{fig:simulation-results} and \crefformat{figure}{#2#1{(c)}#3}\cref{fig:simulation-results} but with the activity variables initialized close to zero. Figure \crefformat{figure}{#2#1{(a)}#3}\cref{fig:random-figure} shows the wavenumber $k_0$, the spatial frequency multiplied to the length of the ring, $L\cdot f_x$, (center row, effectively describing the number of bumps on the ring), and temporal frequency, $f_t$, (bottom row) for the patterns emerging in Turing- and Hopf-unstable states for increasing adaptation strength. We see that further away from the boundaries, inside the regime of Hopf instability, purely temporal oscillations emerge with higher temporal frequencies compared to the activity patterns which emerge after initializing close to the Hopf-unstable fixed point. Closer to the boundaries, purely spatial patterns emerge for $b=0$ that become spatio-temporal for $b\neq0$. Figure \crefformat{figure}{#2#1{(b)}#3}\cref{fig:random-figure} shows the ratio between the dominant spatial frequency $f_x$ and the wavenumbers $k_0$ and $k_{max}$ for $b=-0.5$ plotted against each other (left panel) and the distribution of the differences $k_{max}-k_0$ (right panel). As in the case of Fig. \crefformat{figure}{#2#1{(c)}#3}\cref{fig:simulation-results}, we see that the simulated spatial frequencies predominantly correspond to $k_{max}$ and that $k_0>k_{max}$, except for some locations inside the Hopf instability regime, where the simulated activity patterns were all identified as irregular. 

\subsection{Spatial frequencies of the spatio-temporal patterns emerging for different initializations}\label{appendix:consistency}
\begin{figure}[h!]
    \centering
    \includegraphics[width=0.9\linewidth]{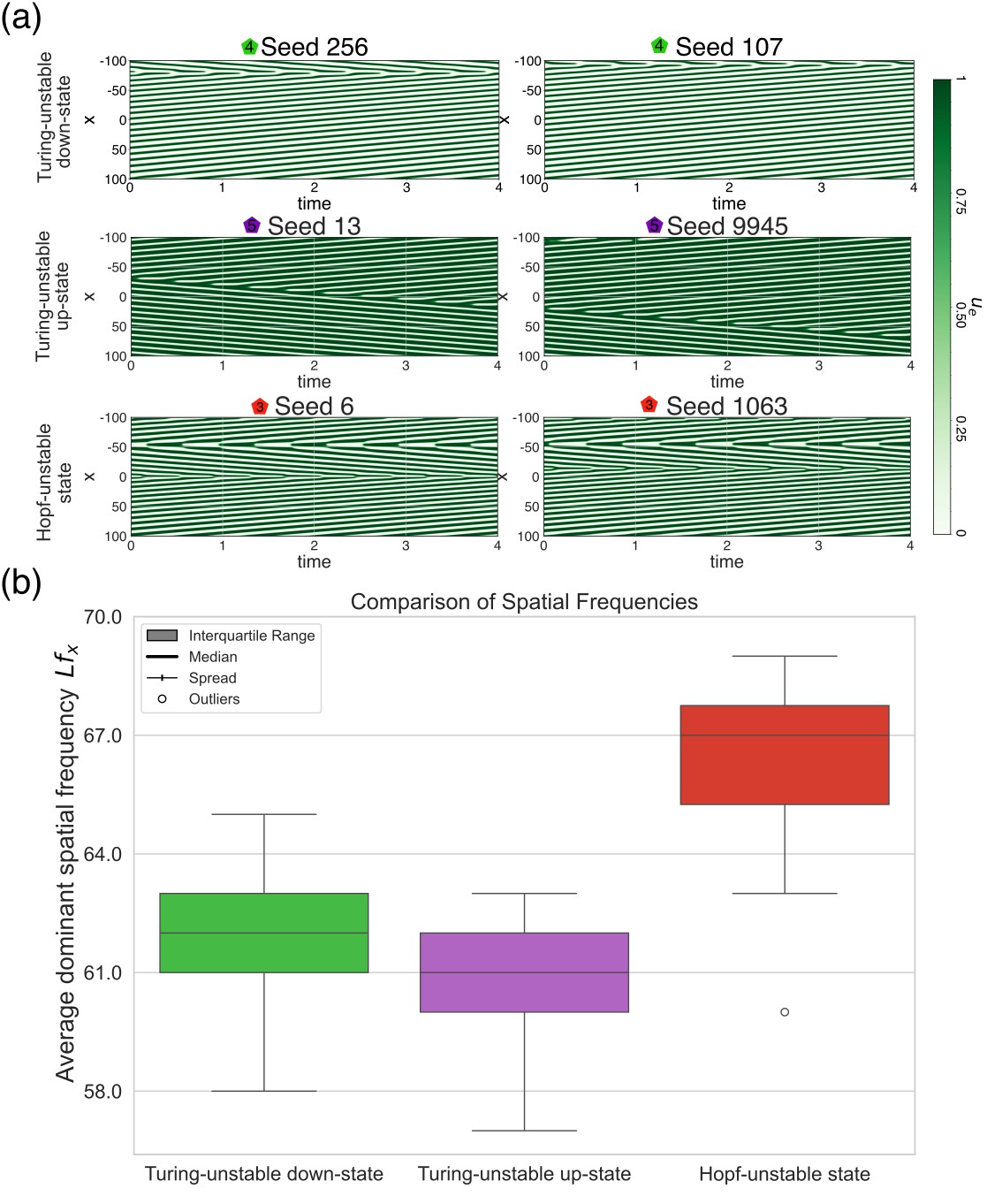}
    \caption{\textbf{(a)} Activity traces 4, 6, and 3 (top to bottom row) of Fig.\ \ref{fig:equivalency-results-with-activity} initialized close to the fixed points of Turing instability in a down-state (first row), in an up-state (second row), and of Hopf instability but with different seeds. \textbf{(b)} Consistency of the spatial dynamics for 30 different seeds for each activity in \rm{(a)}. Visualized are interquartile ranges (boxes), medians (horizontal lines inside boxes), spread (error bars), and outliers outside of spread (circles) for the Turing-unstable down- (green) and up-state (purple), and the Hopf-unstable state (red).}
    \label{fig:consistency}
\end{figure}

\subsection{Input pair tables for Figures \ref{fig:equivalency-results-with-activity} and \ref{fig:speed-results}}\label{appendix:state-space-activity-pairs}\vspace{-0.1cm}
\begin{table}[h!]
    \centering
    \caption{Input pairs $(I_e, I_i)$ for Fig.\ \ref{fig:equivalency-results-with-activity}.}
    \begin{tabular}{|l|c|c|c|c|c|c|c|}
        \hline
        Location  & 0 & 1 & 2 & 3 & 4 & 5 & 6 \\
        \hline
        \hline
        $I_e$ & $-0.65$ & $0$  &  $-0.05$  &  $-0.5$  &  $-0.55$  &  $0.4$  &  $-0.55$ \\
        $I_i$ & $-1$ & $-0.75$  &  $-1.6$  &  $-1.25$  &  $-0.425$  &  $0.2$  &  $-2$  \\
        \hline
    \end{tabular}
    \label{tab:state-space-input-pairs}
\end{table}
\begin{table}[h!]
    \centering
    \caption{Input pairs $(I_e, I_i)$ for Fig.\ \ref{fig:speed-results}.}
    \begin{tabular}{|l|c|c|c|c|c|c|}
        \hline
        Location  & 1 & 2 & 3 & 4 & 5 & 6 \\
        \hline
        \hline
        $I_e$ &  $-0.6$  &  $-0.85$  &  $-0.025$  &  $-0.275$  &  $-1.1$  &  $-1.35$ \\
        $I_i$ &  $-0.875$  &  $-0.875$  &  $-0.2$  &  $-0.2$  &  $-2.325$  &  $-2.325$  \\
        \hline
    \end{tabular}
    \label{tab:speed-input-pairs}
\end{table}


\newpage
\bibliographystyle{apsrev4-2}
\bibliography{refs}

@misc{StroemsdoerferFramework,
  author       = {Ronja Str\"{o}msd\"{o}rfer},
  title        = {Framework for reproducibility},
  howpublished = {\href{https://zenodo.org/records/19455050}{Zenodo, version 1.1}},
  year         = {(2026)},
  doi          = {10.5281/zenodo.19455050},
}

@article{Dimulescu2025,
	author = {Dimulescu, Cristiana and Str{\"o}msd{\"o}rfer, Ronja and Fl{\"o}el, Agnes and Obermayer, Klaus},
	isbn = {2674-0109},
	journal = {Frontiers in Network Physiology},
	n2 = {The human brain is a complex dynamical system which displays a wide range of macroscopic and mesoscopic patterns of neural activity, whose mechanistic origin remains poorly understood. Whole-brain modelling allows us to explore candidate mechanisms causing the observed patterns. However, it is not fully established how the choice of model type and the networks'spatial resolution influence the simulation results, hence, it remains unclear, to which extent conclusions drawn from these results are limited by modelling artefacts. Here, we compare the dynamics of a biophysically realistic, linear-nonlinear cascade model of whole-brain activity with a phenomenological Wilson-Cowan model using three structural connectomes based on the Schaefer parcellation scheme with 100, 200, and 500 nodes. Both neural mass models implement the same mechanistic hypotheses, which specifically address the interaction between excitation, inhibition, and a slow adaptation current which affects the excitatory populations. We quantify the emerging dynamical states in detail and investigate how consistent results are across the different model variants. Then we apply both model types to the specific phenomenon of slow oscillations, which are a prevalent brain rhythm during deep sleep. We investigate the consistency of model predictions when exploring specific mechanistic hypotheses about the effects of both short- and long-range connections and of the antero-posterior structural connectivity gradient on key properties of these oscillations. Overall, our results demonstrate that the coarse-grained dynamics is robust to changes in both model type and network resolution. In some cases, however, model predictions do not generalize. Thus, some care must be taken when interpreting model results.},
	title = {On the robustness of the emergent spatiotemporal dynamics in biophysically realistic and phenomenological whole-brain models at multiple network resolutions},
	type = {Original Research},
	url = {https://www.frontiersin.org/journals/network-physiology/articles/10.3389/fnetp.2025.1589566},
	volume = {5},
	year = {2025}}

@article{Sanchez-Vives2017,
	author = {Sanchez-Vives, Maria V. and Massimini, Marcello and Mattia, Maurizio},
	date = {2017/06/07},
	doi = {10.1016/j.neuron.2017.05.015},
	isbn = {0896-6273},
	journal = {Neuron},
	journal1 = {Neuron},
	month = {2025/08/01},
	number = {5},
	pages = {993--1001},
	publisher = {Elsevier},
	title = {Shaping the Default Activity Pattern of the Cortical Network},
	type = {doi: 10.1016/j.neuron.2017.05.015},
	url = {https://doi.org/10.1016/j.neuron.2017.05.015},
	volume = {94},
	year = {2017},
	year1 = {2017}}

@article{Andola2017,
    author = {D'Andola, Mattia and Rebollo, Beatriz and Casali, Adenauer G and Weinert, Julia F and Pigorini, Andrea and Villa, Rosa and Massimini, Marcello and Sanchez-Vives, Maria V},
    title = {Bistability, Causality, and Complexity in Cortical Networks: An In Vitro Perturbational Study},
    journal = {Cerebral Cortex},
    volume = {28},
    number = {7},
    pages = {2233-2242},
    year = {2017},
    month = {05},issn = {1047-3211},
    doi = {10.1093/cercor/bhx122},
    url = {https://doi.org/10.1093/cercor/bhx122},
}

@article{Sanchez-Vives2020,
    author = {Sanchez-Vives, Maria V},
	date = {2020/06/01/},
	doi = {https://doi.org/10.1016/j.cophys.2020.04.005},
	isbn = {2468-8673},
	journal = {Current Opinion in Physiology},
	pages = {217--223},
	title = {Origin and dynamics of cortical slow oscillations},
	url = {https://www.sciencedirect.com/science/article/pii/S2468867320300365},
	volume = {15},
	year = {2020}}

@article{Steriade1993,
	author = {Steriade, M and Nunez, A and Amzica, F},
	date = {1993/08/01},
	doi = {10.1523/JNEUROSCI.13-08-03252.1993},
	journal = {The Journal of Neuroscience},
	journal1 = {J. Neurosci.},
	lp = {3265},
	month = {08},
	n2 = {We describe a novel slow oscillation in intracellular recordings from cortical association areas 5 and 7, motor areas 4 and 6, and visual areas 17 and 18 of cats under various anesthetics. The recorded neurons (n = 254) were antidromically and orthodromically identified as corticothalamic or callosal elements receiving projections from appropriate thalamic nuclei as well as from homotopic foci in the contralateral cortex. Two major types of cells were recorded: regular- spiking (mainly slow-adapting, but also fast-adapting) neurons and intrinsically bursting cells. A group of slowly oscillating neurons (n = 21) were intracellularly stained and found to be pyramidal-shaped cells in layers III-VI, with luxuriant basal dendritic arbors. The slow rhythm appeared in 88{\%} of recorded neurons. It consisted of slow depolarizing envelopes (lasting for 0.8--1.5 sec) with superimposed full action potentials or presumed dendritic spikes, followed by long- lasting hyperpolarizations. Such sequences recurred rhythmically at less than 1 Hz, with a prevailing oscillation between 0.3 and 0.4 Hz in 67{\%} of urethane-anesthetized animals. While in most neurons (approximately 70{\%}) the repetitive spikes superimposed on the slow depolarization were completely blocked by slight DC hyperpolarization, 30{\%} of cells were found to display relatively small (3--12 mV), rapid, all-or-none potentials after obliteration of full action potentials. These fast spikes were suppressed in an all-or-none fashion at Vm more negative than -90 mV. The depolarizing envelope of the slow rhythm was reduced or suppressed at a Vm of -90 to -100 mV and its duration was greatly reduced by administration of the NMDA blocker ketamine. In keeping with this action, most (56{\%}) neurons recorded in animals under ketamine and nitrous oxide or ketamine and xylazine anesthesia displayed the slow oscillation at higher frequencies (0.6--1 Hz) than under urethane anesthesia (0.3--0.4 Hz). In 18{\%} of the oscillating cells, the slow rhythm mainly consisted of repetitive (15--30 Hz), relatively short-lasting (15--25 msec) IPSPs that could be revealed by bringing the Vm at more positive values than -70 mV. The long-lasting (approximately 1 sec) hyperpolarizing phase of the slow oscillation was best observed at the resting Vm and was reduced at about -100 mV. Simultaneous recording of another cell across the membrane demonstrated synchronous inhibitory periods in both neurons. Intracellular diffusion of Cl- or Cs+ reduced the amplitude and/or duration of cyclic long- lasting hyperpolaryzations.(ABSTRACT TRUNCATED AT 400 WORDS)},
	number = {8},
	pages = {3252},
	title = {A novel slow ($\leq$ 1 {H}z) oscillation of neocortical neurons in vivo: depolarizing and hyperpolarizing components},
	url = {http://www.jneurosci.org/content/13/8/3252.abstract},
	volume = {13},
	year = {1993}}

@article{Loomis1935,
  author    = {Loomis, A. L. and Harvey, E. N. and Hobart, G.},
  title     = {Further Observations on the Potential Rhythms of the Cerebral Cortex During Sleep},
  journal   = {Science},
  year      = {1935},
  volume    = {82},
  number    = {2122},
  pages     = {198--200},
  doi       = {10.1126/science.82.2122.198},
  pmid      = {17844579}
}

@article{Bressloff2012,author = {Paul C Bressloff},
	doi = {10.1088/1751-8113/45/3/033001},
	journal = {Journal of Physics A: Mathematical and Theoretical},
	month = {dec},
	number = {3},
	pages = {033001},
	publisher = {IOP Publishing},
	title = {Spatiotemporal dynamics of continuum neural fields},
	url = {https://dx.doi.org/10.1088/1751-8113/45/3/033001},
	volume = {45},
	year = {2011}}

@article{PintoErmentrout2001,
	author = {Pinto, David J. and Ermentrout, G. Bard},
	date = {2001/01/01},
	doi = {10.1137/S0036139900346453},
	isbn = {0036-1399},
	journal = {SIAM Journal on Applied Mathematics},
	journal1 = {SIAM Journal on Applied Mathematics},
	journal2 = {SIAM J. Appl. Math.},
	month = {2025/09/22},
	n2 = {We consider traveling front and pulse solutions to a system of integro-differential equations used to describe the activity of synaptically coupled neuronal networks in a single spatial dimension. Our first goal is to establish a series of direct links between the abstract nature of the equations and their interpretation in terms of experimental findings in the cortex and other brain regions. This is accomplished first by presenting a biophysically motivated derivation of the system and then by establishing a framework for comparison between numerical and experimental measures of activity propagation speed. Our second goal is to establish the existence of traveling pulse solutions using more rigorous methods. Two techniques are presented. The first, a shooting argument, reduces the problem from finding a specific solution to an integro-differential equation system to finding any solution to an ODE system. The second, a singular perturbation argument, provides a construction of traveling pulse solutions under more general conditions.},
	number = {1},
	pages = {206--225},
	publisher = {Society for Industrial and Applied Mathematics},
	title = {Spatially Structured Activity in Synaptically Coupled Neuronal Networks: I. Traveling Fronts and Pulses},
	type = {doi: 10.1137/S0036139900346453},
	url = {https://doi.org/10.1137/S0036139900346453},
	volume = {62},
	year = {2001},
	year1 = {2001}}

@article{Dasilva2012,author = {Dasilva, Miguel and Camassa, Alessandra and Navarro-Guzman, Alvaro and Pazienti, Antonio and Perez-Mendez, Lorena and Zamora-L{\'o}pez, Gorka and Mattia, Maurizio and Sanchez-Vives, Maria V.},
	date = {2021/01/01/},
	doi = {https://doi.org/10.1016/j.neuroimage.2020.117415},
	isbn = {1053-8119},
	journal = {NeuroImage},
	pages = {117415},
	title = {Modulation of cortical slow oscillations and complexity across anesthesia levels},
	url = {https://www.sciencedirect.com/science/article/pii/S1053811920309009},
	volume = {224},
	year = {2021}}

@article{Rubino2006,author = {Rubino, Doug and Robbins, Kay A and Hatsopoulos, Nicholas G},
	date = {2006/12/01},
	doi = {10.1038/nn1802},
	id = {Rubino2006},
	isbn = {1546-1726},
	journal = {Nature Neuroscience},
	number = {12},
	pages = {1549--1557},
	title = {Propagating waves mediate information transfer in the motor cortex},
	url = {https://doi.org/10.1038/nn1802},
	volume = {9},
	year = {2006}}

@article{Cakan2022,author = {Caglar Cakan and Cristiana Dimulescu and Liliia Khakimova and Daniela Obst and Agnes Flöel and Klaus Obermayer},
   doi = {10.3389/fncom.2021.800101},
   issn = {16625188},
   journal = {Frontiers in Computational Neuroscience},
   month = {1},
   publisher = {Frontiers Media S.A.},
   title = {Spatiotemporal Patterns of Adaptation-Induced Slow Oscillations in a Whole-Brain Model of Slow-Wave Sleep},
   volume = {15},
   year = {2022},
}

@article{Ganguly2024,author = {Ganguly, Chittotosh and Bezugam, Sai Sukruth and Abs, Elisabeth and Payvand, Melika and Dey, Sounak and Suri, Manan},
	date = {2024/02/01},
	doi = {10.1038/s44172-024-00165-9},
	id = {Ganguly2024},
	isbn = {2731-3395},
	journal = {Communications Engineering},
	number = {1},
	pages = {22},
	title = {Spike frequency adaptation: bridging neural models and neuromorphic applications},
	url = {https://doi.org/10.1038/s44172-024-00165-9},
	volume = {3},
	year = {2024}}

@article{Harris2018,author = {Jeremy D. Harris and Bard Ermentrout},
   doi = {10.1016/j.physd.2017.12.011},
   issn = {01672789},
   journal = {Physica D: Nonlinear Phenomena},
   month = {4},
   pages = {30-46},
   publisher = {Elsevier B.V.},
   title = {Traveling waves in a spatially-distributed {W}ilson–{C}owan model of cortex: From fronts to pulses},
   volume = {369},
   year = {2018},
}

@article{Hong-Yuan2010,author = {Chu, Hong-yuan and Zhen, Xuechu},
	date = {2010/09/01},
	doi = {10.1038/aps.2010.105},
	id = {Chu2010},
	isbn = {1745-7254},
	journal = {Acta Pharmacologica Sinica},
	number = {9},
	pages = {1036--1043},
	title = {Hyperpolarization-activated, cyclic nucleotide-gated (HCN) channels in the regulation of midbrain dopamine systems},
	url = {https://doi.org/10.1038/aps.2010.105},
	volume = {31},
	year = {2010}}

@article{Jercog2017,
author = {Jercog, Daniel and Roxin, Alex and Barth{\'o}, Peter and Luczak, Artur and Compte, Albert and de la Rocha, Jaime},
	c1 = {eLife 2017;6:e22425},
	date = {2017/08/04},
	doi = {10.7554/eLife.22425},
	isbn = {2050-084X},
	journal = {eLife},
	pages = {e22425},
	publisher = {eLife Sciences Publications, Ltd},
	title = {UP-DOWN cortical dynamics reflect state transitions in a bistable network},
	url = {https://doi.org/10.7554/eLife.22425},
	volume = {6},
	year = {2017}}

@article{Levenstein2019,author = {Levenstein, Daniel and Buzs{\'a}ki, Gy{\"o}rgy and Rinzel, John},
	date = {2019/06/06},
	doi = {10.1038/s41467-019-10327-5},
	id = {Levenstein2019},
	isbn = {2041-1723},
	journal = {Nature Communications},
	number = {1},
	pages = {2478},
	title = {NREM sleep in the rodent neocortex and hippocampus reflects excitable dynamics},
	url = {https://doi.org/10.1038/s41467-019-10327-5},
	volume = {10},
	year = {2019}}

@article{Massimini2004,
	author = {Massimini, Marcello and Huber, Reto and Ferrarelli, Fabio and Hill, Sean and Tononi, Giulio},
	date = {2004/08/04},
	doi = {10.1523/JNEUROSCI.1318-04.2004},
	journal = {The Journal of Neuroscience},
	journal1 = {J. Neurosci.},
	lp = {6870},
	month = {08},
	n2 = {During much of sleep, virtually all cortical neurons undergo a slow oscillation (\&lt;1 Hz) in membrane potential, cycling from a hyperpolarized state of silence to a depolarized state of intense firing. This slow oscillation is the fundamental cellular phenomenon that organizes other sleep rhythms such as spindles and slow waves. Using high-density electroencephalogram recordings in humans, we show here that each cycle of the slow oscillation is a traveling wave. Each wave originates at a definite site and travels over the scalp at an estimated speed of 1.2-7.0 m/sec. Waves originate more frequently in prefrontal-orbitofrontal regions and propagate in an anteroposterior direction. Their rate of occurrence increases progressively reaching almost once per second as sleep deepens. The pattern of origin and propagation of sleep slow oscillations is reproducible across nights and subjects and provides a blueprint of cortical excitability and connectivity. The orderly propagation of correlated activity along connected pathways may play a role in spike timing-dependent synaptic plasticity during sleep.},
	number = {31},
	pages = {6862},
	title = {The Sleep Slow Oscillation as a Traveling Wave},
	url = {http://www.jneurosci.org/content/24/31/6862.abstract},
	volume = {24},
	year = {2004}}

@article{Mattia2012,author = {Mattia, Maurizio and Sanchez-Vives, Maria V.},
	date = {2012/06/01},
	doi = {10.1007/s11571-011-9179-4},
	id = {Mattia2012},
	isbn = {1871-4099},
	journal = {Cognitive Neurodynamics},
	number = {3},
	pages = {239--250},
	title = {Exploring the spectrum of dynamical regimes and timescales in spontaneous cortical activity},
	url = {https://doi.org/10.1007/s11571-011-9179-4},
	volume = {6},
	year = {2012}}

@article{McCormick1990,address = {Section of Neuroanatomy, Yale University School of Medicine, New Haven, CT 06510.},
	author = {McCormick, D A and Pape, H C},
	crdt = {1990/12/01 00:00},
	date = {1990 Dec},
	dcom = {19910822},
	doi = {10.1113/jphysiol.1990.sp018331},
	edat = {1990/12/01 00:00},
	issn = {0022-3751 (Print); 1469-7793 (Electronic); 0022-3751 (Linking)},
	jid = {0266262},
	journal = {J Physiol},
	jt = {The Journal of physiology},
	language = {eng},
	lr = {20220310},
	mh = {Action Potentials/drug effects/physiology; Animals; Cations/pharmacology; Cats; Chlorides/physiology; Female; Guinea Pigs; In Vitro Techniques; Ion Channels/*physiology; Male; Membrane Potentials/physiology; Neurons/*physiology; Potassium/physiology; Sodium/physiology; Thalamic Nuclei/physiology; Thalamus/*physiology; Time Factors},
	mhda = {1990/12/01 00:01},
	month = {Dec},
	own = {NLM},
	pages = {291--318},
	phst = {1990/12/01 00:00 {$[$}pubmed{$]$}; 1990/12/01 00:01 {$[$}medline{$]$}; 1990/12/01 00:00 {$[$}entrez{$]$}; 1990/12/01 00:00 {$[$}pmc-release{$]$}},
	pl = {England},
	pmc = {PMC1181775},
	pmcr = {1990/12/01},
	pmid = {1712843},
	pst = {ppublish},
	pt = {Journal Article; Research Support, Non-U.S. Gov't; Research Support, U.S. Gov't, P.H.S.},
	rn = {0 (Cations); 0 (Chlorides); 0 (Ion Channels); 9NEZ333N27 (Sodium); RWP5GA015D (Potassium)},
	sb = {IM},
	status = {MEDLINE},
	title = {Properties of a hyperpolarization-activated cation current and its role in rhythmic oscillation in thalamic relay neurones.},
	volume = {431},
	year = {1990}}

@article{Mehrotra2024,
	author = {Mehrotra, Dhruv and Levenstein, Daniel and Duszkiewicz, Adrian J. and Carrasco, Sofia Skromne and Booker, Sam A. and Kwiatkowska, Angelika and Peyrache, Adrien},
	date = {2024/07/22/},
	doi = {https://doi.org/10.1016/j.cub.2024.05.048},
	isbn = {0960-9822},
	journal = {Current Biology},
	number = {14},
	pages = {3043--3054.e8},
	title = {Hyperpolarization-activated currents drive neuronal activation sequences in sleep},
	url = {https://www.sciencedirect.com/science/article/pii/S0960982224006912},
	volume = {34},
	year = {2024}}

@article{Porta2024,
	author = {Dalla Porta, Leonardo and Barbero-Castillo, Almudena and Sanchez-Sanchez, Jos{\'e}Manuel and Cancino, Nathalia and Sanchez-Vives, Maria V.},
	date = {2025/04/01},
	doi = {https://doi.org/10.1113/JP287616},
	isbn = {0022-3751},
	journal = {The Journal of Physiology},
	journal1 = {The Journal of Physiology},
	journal2 = {The Journal of Physiology},
	journal3 = {J Physiol},
	month = {2025/10/06},
	n2 = {Abstract Understanding the link between cellular processes and brain function remains a key challenge in neuroscience. One crucial aspect is the interplay between specific ion channels and network dynamics. This work reveals a role for h-current, a hyperpolarization-activated cationic current, in shaping cortical slow oscillations. Cortical slow oscillations are generated not only during slow wave sleep and deep anaesthesia, but also in association with disorders of consciousness and brain lesions. Cortical slow oscillations exhibit rhythmic periods of activity (Up states) alternating with silent periods (Down states). By progressively reducing h-current in both cortical slices and in a computational model, we observed Up states transformed into prolonged plateaus of sustained firing, while Down states were also significantly extended. This transformation led to a fivefold reduction in oscillation frequency. In a biophysical recurrent network model, we identified the cellular mechanisms underlying this transformation of network dynamics: an increased neuronal input resistance and membrane time constant, increasing neuronal responsiveness to even weak inputs. A partial block of h-current therefore resulted in a change in brain state. HCN (hyperpolarization-activated cyclic nucleotide-gated) channels, which generate h-current, are known targets for neuromodulation, suggesting potential pathways for dynamic control of brain rhythms. Key points We investigated the role of h-current in shaping emergent cortical slow oscillation dynamics, specifically Up and Down states, in cortical slices. Blocking h-current transformed Up states into prolonged plateaus of sustained firing, lasting up to 4 s. Down states were also significantly elongated and the oscillatory frequency decreased. A biophysical model of the cortical network replicated these findings and allowed us to explore the underlying mechanisms. An increase in cellular input resistance and time constant led to a rise in network excitability, synaptic responsiveness and firing rates. Our results highlight the significant role of h-current in controlling cortical slow rhythmic patterns, making it a relevant target for neuromodulators regulating brain states.},
	number = {8},
	pages = {2409--2424},
	publisher = {John Wiley \& Sons, Ltd},
	title = {H-current modulation of cortical Up and Down states},
	url = {https://doi.org/10.1113/JP287616},
	volume = {603},
	year = {2025},
	year1 = {2025}}

@article{RaschBorn2013,
    author = {Rasch, Bj\"{o}rn and Born, Jan},
	doi = {10.1152/physrev.00032.2012},
	journal = {Physiological Reviews},
	number = {2},
	pages = {681-766},
	title = {About Sleep's Role in Memory},
	url = {https://doi.org/10.1152/physrev.00032.2012},
	volume = {93},
	year = {2013}}

@article{Klinzing2019,author = {Klinzing, Jens G. and Niethard, Niels and Born, Jan},
	date = {2019/10/01},
	doi = {10.1038/s41593-019-0467-3},
	id = {Klinzing2019},
	isbn = {1546-1726},
	journal = {Nature Neuroscience},
	number = {10},
	pages = {1598--1610},
	title = {Mechanisms of systems memory consolidation during sleep},
	url = {https://doi.org/10.1038/s41593-019-0467-3},
	volume = {22},
	year = {2019}}

@article{Brodt2023,author = {Brodt, Svenja and Inostroza, Marion and Niethard, Niels and Born, Jan},
	date = {2023/04/05/},
	doi = {https://doi.org/10.1016/j.neuron.2023.03.005},
	isbn = {0896-6273},
	journal = {Neuron},
	number = {7},
	pages = {1050--1075},
	title = {Sleep---A brain-state serving systems memory consolidation},
	url = {https://www.sciencedirect.com/science/article/pii/S0896627323002015},
	volume = {111},
	year = {2023}}

@article{Ha2017,
title = "Spike frequency adaptation in neurons of the central nervous system",
author = "Ha, Go Eun and Eunji Cheong",
year = "2017",
month = aug,
day = "1",
doi = "10.5607/en.2017.26.4.179",
language = "English",
volume = "26",
pages = "179--185",
journal = "Experimental Neurobiology",
issn = "1226-2560",
publisher = "Korean Society for Neurodegenerative Disease",
number = "4",
}

@article{BrownAdams1980,
	author = {Brown, D. A. and Adams, P. R.},
	date = {1980/02/01},
	doi = {10.1038/283673a0},
	id = {Brown1980},
	isbn = {1476-4687},
	journal = {Nature},
	number = {5748},
	pages = {673--676},
	title = {Muscarinic suppression of a novel voltage-sensitive K+ current in a vertebrate neurone},
	url = {https://doi.org/10.1038/283673a0},
	volume = {283},
	year = {1980}}

@article{Schwindt1989,
	author = {Schwindt, P. C. and Spain, W. J. and Crill, W. E.},
	date = {1989/02/01},
	doi = {10.1152/jn.1989.61.2.233},
	isbn = {0022-3077},
	journal = {Journal of Neurophysiology},
	journal1 = {Journal of Neurophysiology},
	month = {2025/10/02},
	n2 = {1. The function and ionic mechanism of a slow outward current were studied in large layer V neurons of cat sensorimotor cortex using an in vitro slice preparation and single microelectrode voltage clamp. 2. With Ca2+ influx blocked, a slow relaxation ("tail") of outward current followed either (1) repetitive firing evoked for 1 s or (2) a small 1-s depolarizing voltage clamp step that activated the persistent Na+ current of neocortical neurons, INaP. When a depolarization that activated INaP was maintained, an outward current gradually developed and increased in amplitude over a period of tens of seconds to several minutes. An outward tail current of similar duration followed repolarization. The slow outward current was abolished by TTX, indicating it depended on Na+ influx. 3. With Ca2+ influx blocked, the onset of the slow Na+-dependent outward current caused spike frequency adaptation during current-evoked repetitive firing. Following the firing, the decay of the Na+-dependent current caused a slow afterhyperpolarization (sAHP) and a long-lasting reduction of excitability. It also was responsible for habituation of the response to repeated identical current pulses. 4. The Na+-dependent tail current had properties expected of a K+ current. Membrane chord conductance increased during the tail, and tail amplitude was reduced or reversed by membrane potential hyperpolarization and raised extracellular K+ concentration {$[$}( K+{$]$}0). 5. The current tail was reduced reversibly by the K+ channel blockers TEA (5-10 mM), muscarine (5-20 microM), and norepinephrine (100 microM). These agents also resulted in a larger, more sustained inward current during the preceding step depolarization. Comparison of current time course before and after the application of blocking agents suggested that, in spite of its capability for slow buildup and decay, the onset of the Na+-dependent outward current occurs within 100 ms of an adequate step depolarization. 6. With Ca2+ influx blocked, extracellular application of dantrolene sodium (30 microM) had no clear effect on the current tail or the corresponding sAHP.(ABSTRACT TRUNCATED AT 400 WORDS)},
	number = {2},
	pages = {233--244},
	publisher = {American Physiological Society},
	title = {Long-lasting reduction of excitability by a sodium-dependent potassium current in cat neocortical neurons},
	type = {doi: 10.1152/jn.1989.61.2.233},
	url = {https://doi.org/10.1152/jn.1989.61.2.233},
	volume = {61},
	year = {1989},
	year1 = {1989}}

@article{BrownGriffith1983,
	author = {Brown, D A and Griffith, W H},
	date = {1983/04/01},
	doi = {https://doi.org/10.1113/jphysiol.1983.sp014624},
	isbn = {0022-3751},
	journal = {The Journal of Physiology},
	journal1 = {The Journal of Physiology},
	journal2 = {The Journal of Physiology},
	month = {2025/10/02},
	n2 = {Slow clamp currents were recorded from CA1 and CA3 pyramidal neurones in slices of guinea-pig hippocampus maintained in vitro, using a single micro-electrode sample-and-hold technique. Depolarizing voltage commands evoked a time- and voltage-dependent outward current which was suppressed by removing external Ca or by adding Cd (0.5 mM) or Mn (5 mM). This Ca-dependent current (Ic) was not reduced by muscarinic agonists (unlike IM) but was greatly reduced by 5-20 mM-tetraethylammonium (TEA). Repolarizing IC tail currents reversed at -73 +/- 5 mV in 3 mM-K solution. The reversal potential became about 30 mV more positive on raising {$[$}K{$]$}o to 15 mM. No clear change in current amplitude or tail-current reversal potential occurred on adding Cs (2 mM), reducing {$[$}Cl{$]$}o from 128 to 10 mM, or replacing external Na with Tris. The underlying conductance GC was activated at membrane potentials positive to -45 mV. At -32 mV GC showed an approximately exponential increase with time, with a time constant of approximately 0.6 sec at 26 degrees C. Repolarizing tail currents declined exponentially with time, the time constant becoming shorter with increasing negative post-pulse potentials. When the clamp was switched off at the end of a depolarizing command of sufficient amplitude and duration to activate IC, a membrane hyperpolarization to -73 mV ensued, of similar amplitude and decay time to that following spontaneous action potentials. It is concluded that the clamp current observed in these experiments is probably the Ca-activated K current thought to contribute to the post-activation after-hyperpolarization in hippocampal neurones.},
	number = {1},
	pages = {287--301},
	publisher = {John Wiley \& Sons, Ltd},
	title = {Calcium-activated outward current in voltage-clamped hippocampal neurones of the guinea-pig.},
	url = {https://doi.org/10.1113/jphysiol.1983.sp014624},
	volume = {337},
	year = {1983},
	year1 = {1983}}

@article{Upchurch2022,
  author    = {Upchurch, C. M. and Combe, C. L. and Knowlton, C. J. and Rousseau, V. G. and Gasparini, S. and Canavier, C. C.},
  title     = {Long-Term Inactivation of Sodium Channels as a Mechanism of Adaptation in CA1 Pyramidal Neurons},
  journal   = {Journal of Neuroscience},
  year      = {2022},
  volume    = {42},
  number    = {18},
  pages     = {3768--3782},
  doi       = {10.1523/JNEUROSCI.1914-21.2022},
  pmid      = {35332085},
  pmcid     = {PMC9087813},
  note      = {Epub 2022 Mar 24}
}

@article{Faber2003,
	author = {Faber, E. S. Louise and Sah, Pankaj},
	date = {2003/06/01},
	doi = {10.1177/1073858403009003011},
	isbn = {1073-8584},
	journal = {The Neuroscientist},
	journal1 = {The Neuroscientist},
	journal2 = {Neuroscientist},
	month = {2025/09/15},
	n2 = {Calcium-activated potassium channels are a large family of potassium channels that are found throughout the central nervous system and in many other cell types. These channels are activated by rises in cytosolic calcium largely in response to calcium influx via voltage-gated calcium channels that open during action potentials. Activation of these potassium channels is involved in the control of a number of physiological processes from the firing properties of neurons to the control of transmitter release. These channels form the target for modulation for a range of neurotransmitters and have been implicated in the pathogenesis of neurological and psychiatric disorders. Here the authors summarize the varieties of calcium-activated potassium channels present in central neurons and their defining molecular and biophysical properties.},
	number = {3},
	pages = {181--194},
	publisher = {SAGE Publications Inc STM},
	title = {Calcium-Activated Potassium Channels: Multiple Contributions to Neuronal Function},
	type = {doi: 10.1177/1073858403009003011},
	url = {https://doi.org/10.1177/1073858403009003011},
	volume = {9},
	year = {2003},
	year1 = {2003}}

@article{SanchezVivez2020,
    author = {Maria V Sanchez-Vives},
	doi = {https://doi.org/10.1016/j.cophys.2020.04.005},
	issn = {2468-8673},
	journal = {Current Opinion in Physiology},
	pages = {217-223},
	title = {Origin and dynamics of cortical slow oscillations},
	url = {https://www.sciencedirect.com/science/article/pii/S2468867320300365},
	volume = {15},
	year = {2020}}

@article{SanchezVivez2000,author = {Sanchez-Vives, Maria V. and McCormick, David A.},
	date = {2000/10/01},
	doi = {10.1038/79848},
	id = {Sanchez-Vives2000},
	isbn = {1546-1726},
	journal = {Nature Neuroscience},
	number = {10},
	pages = {1027--1034},
	title = {Cellular and network mechanisms of rhythmic recurrent activity in neocortex},
	url = {https://doi.org/10.1038/79848},
	volume = {3},
	year = {2000}}

@article{Nghiem2020,
  author    = {Nghiem, Trang-Anh E. and Tort-Colet, Núria and G{\'o}rski, Tomasz and Ferrari, Ulisse and Moghimyfiroozabad, Shayan and Goldman, Jennifer S. and Tele{\'n}czuk, Bartosz and Capone, Cristiano and Bal, Thierry and di Volo, Matteo and Destexhe, Alain},
  title     = {Cholinergic Switch between Two Types of Slow Waves in Cerebral Cortex},
  journal   = {Cerebral Cortex},
  year      = {2020},
  volume    = {30},
  number    = {6},
  pages     = {3451--3466},
  doi       = {10.1093/cercor/bhz320},
  issn      = {1047-3211},
  url       = {https://doi.org/10.1093/cercor/bhz320}
}

@article{Shu2023,author = {Shu, Yousheng and Hasenstaub, Andrea and McCormick, David A.},
	doi = {10.1101/2023.07.12.548753},
	elocation-id = {2023.07.12.548753},
	journal = {bioRxiv},
	publisher = {Cold Spring Harbor Laboratory},
	title = {The h-current controls cortical recurrent network activity through modulation of dendrosomatic communication},
	url = {https://www.biorxiv.org/content/early/2023/07/13/2023.07.12.548753},
	year = {2023}}

@article{Roth2020,author = {Roth, Fabian C. and Hu, Hua},
	date = {2020/05/07},
	doi = {10.1038/s41467-020-15791-y},
	id = {Roth2020},
	isbn = {2041-1723},
	journal = {Nature Communications},
	number = {1},
	pages = {2248},
	title = {An axon-specific expression of HCN channels catalyzes fast action potential signaling in GABAergic interneurons},
	url = {https://doi.org/10.1038/s41467-020-15791-y},
	volume = {11},
	year = {2020}}

@article{Rich2025,
	author = {Rich, Scott and Valiante, Taufik A. and Lefebvre, J{\'e}r{\'e}mie},
	date = {2025/06/30},
	journal = {PLOS Computational Biology},
	journal1 = {PLOS Computational Biology},
	month = {06},
	n2 = {Author summary How a vast array of insults to the healthy brain can all lead to the same pathological endpoint of seizure remains one of the most perplexing questions in epilepsy research. Historically, computational modeling has been successfully used to derive mechanisms relating many such insults to seizure onset. These pathways are often classified by how particular insults perturb the balance between excitation and inhibition in neuronal microcircuits; as an example, underexpression of both the h- and m-channels are thought to lead to seizure by promoting excessive excitation or insufficient inhibition. This intuition, though, is challenged by the seeming paradox that the overexpression of both these ion channels is also observed in some epilepsies. Here, we address this lingering question using in silico approaches by first creating a model that produces seizure-like events dependent upon approximations of the effects of excessive h- and m-channel expression. Analyzing this model leads to a new hypothesis for how these pathologies promote seizure without contradicting previous findings on channel underexpression: such changes instead alter the network's response to unusually correlated inputs. This mechanism recontextualizes our understanding of this relatively understudied cause of epilepsy, providing a host of experimentally testable predictions for future study.},
	number = {6},
	pages = {e1013199--},
	publisher = {Public Library of Science},
	title = {H- and m-channel overexpression promotes seizure-like events by impairing the ability of inhibitory neurons to process correlated inputs},
	type = {doi:10.1371/journal.pcbi.1013199},
	url = {https://doi.org/10.1371/journal.pcbi.1013199},
	volume = {21},
	year = {2025}}

@article{Timofeev2002,
 ISSN = {00278424},
 URL = {http://www.jstor.org/stable/3059233},author = {Igor Timofeev and Maxim Bazhenov and Terrence Sejnowski and Mircea Steriade},
 journal = {Proceedings of the National Academy of Sciences of the United States of America},
 number = {14},
 pages = {9533--9537},
 publisher = {National Academy of Sciences},
 title = {Cortical Hyperpolarization-Activated Depolarizing Current Takes Part in the Generation of Focal Paroxysmal Activities},
 urldate = {2025-03-13},
 volume = {99},
 year = {2002}
}

@article{Papadopoulos2020,author = {Papadopoulos, Lia AND Lynn, Christopher W. AND Battaglia, Demian AND Bassett, Danielle S.},
	doi = {10.1371/journal.pcbi.1008144},
	journal = {PLOS Computational Biology},
	month = {09},
	number = {9},
	pages = {1-43},
	publisher = {Public Library of Science},
	title = {Relations between large-scale brain connectivity and effects of regional stimulation depend on collective dynamical state},
	url = {https://doi.org/10.1371/journal.pcbi.1008144},
	volume = {16},
	year = {2020}}

@article{Torao2021,author = {Torao-Angosto, Melody and Manasanch, Arnau and Mattia, Maurizio and Sanchez-Vives, Maria V.},
	doi = {10.3389/fnsys.2021.609645},
	issn = {1662-5137},
	journal = {Frontiers in Systems Neuroscience},
	title = {Up and Down States During Slow Oscillations in Slow-Wave Sleep and Different Levels of Anesthesia},
	url = {https://www.frontiersin.org/articles/10.3389/fnsys.2021.609645},
	volume = {15},
	year = {2021}}

@article{Byrne2019,
	author = {Byrne, {\'A}ine and Avitabile, Daniele and Coombes, Stephen},
	date = {2019/01/07/},
	day = {07},
	doi = {10.1103/PhysRevE.99.012313},
	id = {10.1103/PhysRevE.99.012313},
	j1 = {PRE},
	journal = {Physical Review E},
	journal1 = {Phys. Rev. E},
	month = {01},
	number = {1},
	pages = {012313--},
	publisher = {American Physical Society},
	title = {Next-generation neural field model: The evolution of synchrony within patterns and waves},
	url = {https://link.aps.org/doi/10.1103/PhysRevE.99.012313},
	volume = {99},
	year = {2019}}

@article{Coombes2005,
	author = {Coombes, S. },
	date = {2005/08/01},
	doi = {10.1007/s00422-005-0574-y},
	id = {Coombes2005},
	isbn = {1432-0770},
	journal = {Biological Cybernetics},
	number = {2},
	pages = {91--108},
	title = {Waves, bumps, and patterns in neural field theories},
	url = {https://doi.org/10.1007/s00422-005-0574-y},
	volume = {93},
	year = {2005}}

@article{Curtu2004,
    author = {Curtu, Rodica and Ermentrout, Bard},
	doi = {10.1137/030600503},
	journal = {SIAM Journal on Applied Dynamical Systems},
	number = {3},
	pages = {191-231},
	title = {Pattern Formation in a Network of Excitatory and Inhibitory Cells with Adaptation},
	url = {https://doi.org/10.1137/030600503},
	volume = {3},
	year = {2004}}

@book{Ermentrout2010mathematical_book,
  title={Mathematical Foundations of Neuroscience},
  author={Ermentrout, G.B. and Terman, D.H.},
  isbn={9780387877075},
  lccn={2010929771},
  series={Interdisciplinary Applied Mathematics},
  url={https://books.google.de/books?id=ZkWiVu\_OXgIC},
  year={2010},
  publisher={Springer New York}
}

@article{Friedlander2009,address = {Departments of Physics, Laboratory of Network Biology Research, Technion - Israel Institute of Technology, Haifa 32000, Israel.},
	author = {Friedlander, Tamar and Brenner, Naama},
	cois = {The authors declare no conflict of interest.},
	crdt = {2009/12/19 06:00},
	date = {2009 Dec 29},
	dcom = {20100217},
	dep = {20091215},
	doi = {10.1073/pnas.0902146106},
	edat = {2009/12/19 06:00},
	issn = {1091-6490 (Electronic); 0027-8424 (Print); 0027-8424 (Linking)},
	jid = {7505876},
	journal = {Proc Natl Acad Sci U S A},
	jt = {Proceedings of the National Academy of Sciences of the United States of America},
	language = {eng},
	lid = {10.1073/pnas.0902146106 {$[$}doi{$]$}},
	lr = {20211020},
	mh = {*Adaptation, Physiological; Feedback, Physiological; Kinetics; Membrane Proteins/antagonists \& inhibitors/chemistry/*metabolism; *Models, Biological; Systems Biology},
	mhda = {2010/02/18 06:00},
	month = {Dec},
	number = {52},
	own = {NLM},
	pages = {22558--22563},
	phst = {2009/12/19 06:00 {$[$}entrez{$]$}; 2009/12/19 06:00 {$[$}pubmed{$]$}; 2010/02/18 06:00 {$[$}medline{$]$}; 2010/06/29 00:00 {$[$}pmc-release{$]$}},
	pii = {0902146106; 0666},
	pl = {United States},
	pmc = {PMC2799740},
	pmcr = {2010/06/29},
	pmid = {20018770},
	pst = {ppublish},
	pt = {Journal Article; Research Support, Non-U.S. Gov't},
	rn = {0 (Membrane Proteins)},
	sb = {IM},
	status = {MEDLINE},
	title = {Adaptive response by state-dependent inactivation.},
	volume = {106},
	year = {2009}}

@inbook{Kilpatrick2013WC,
	address = {New York, NY},
	author = {Kilpatrick, Zachary P.},
	booktitle = {Encyclopedia of Computational Neuroscience},
	doi = {10.1007/978-1-4614-7320-6_80-1},
	editor = {Jaeger, Dieter and Jung, Ranu},
	isbn = {978-1-4614-7320-6},
	pages = {1--5},
	publisher = {Springer New York},
	title = {Wilson-{C}owan Model},
	url = {https://doi.org/10.1007/978-1-4614-7320-6_80-1},
	year = {2013}}

@article{Ladenbauer2014,
    author = {Ladenbauer, Josef and Augustin, Moritz and Obermayer, Klaus},
	doi = {10.1152/jn.00586.2013},
	journal = {Journal of Neurophysiology},
	number = {5},
	pages = {939-953},
	title = {How adaptation currents change threshold, gain, and variability of neuronal spiking},
	url = {https://doi.org/10.1152/jn.00586.2013},
	volume = {111},
	year = {2014}}

@article{Augustin2013,
	author = {Augustin, Moritz and Ladenbauer, Josef and Obermayer, Klaus},
	isbn = {1662-5188},
	journal = {Frontiers in Computational Neuroscience},
	n2 = {<p>Neural mass signals from <italic>in-vivo</italic> recordings often show oscillations with frequencies ranging from {$<$}1 to 100 {H}z. Fast rhythmic activity in the beta and gamma range can be generated by network-based mechanisms such as recurrent synaptic excitation-inhibition loops. Slower oscillations might instead depend on neuronal adaptation currents whose timescales range from tens of milliseconds to seconds. Here we investigate how the dynamics of such adaptation currents contribute to spike rate oscillations and resonance properties in recurrent networks of excitatory and inhibitory neurons. Based on a network of sparsely coupled spiking model neurons with two types of adaptation current and conductance-based synapses with heterogeneous strengths and delays we use a mean-field approach to analyze oscillatory network activity. For constant external input, we find that spike-triggered adaptation currents provide a mechanism to generate slow oscillations over a wide range of adaptation timescales as long as recurrent synaptic excitation is sufficiently strong. Faster rhythms occur when recurrent inhibition is slower than excitation and oscillation frequency increases with the strength of inhibition. Adaptation facilitates such network-based oscillations for fast synaptic inhibition and leads to decreased frequencies. For oscillatory external input, adaptation currents amplify a narrow band of frequencies and cause phase advances for low frequencies in addition to phase delays at higher frequencies. Our results therefore identify the different key roles of neuronal adaptation dynamics for rhythmogenesis and selective signal propagation in recurrent networks.</p>},
	title = {How adaptation shapes spike rate oscillations in recurrent neuronal networks},
	type = {Original Research},
	url = {https://www.frontiersin.org/journals/computational-neuroscience/articles/10.3389/fncom.2013.00009},
	volume = {7},
	year = {2013}}

@article{Meijer2014,author = {Meijer, Hil G. E. and Coombes, Stephen},
	date = {2014/04/01},
	doi = {10.1007/s00285-013-0670-x},
	id = {Meijer2014},
	isbn = {1432-1416},
	journal = {Journal of Mathematical Biology},
	number = {5},
	pages = {1249--1268},
	title = {Travelling waves in a neural field model with refractoriness},
	url = {https://doi.org/10.1007/s00285-013-0670-x},
	volume = {68},
	year = {2014}}

@article{WilsonCowan1972,
	author = {Wilson, Hugh R. and Cowan, Jack D.},
	date = {1972/01/01/},
	doi = {https://doi.org/10.1016/S0006-3495(72)86068-5},
	isbn = {0006-3495},
	journal = {Biophysical Journal},
	number = {1},
	pages = {1--24},
	title = {Excitatory and Inhibitory Interactions in Localized Populations of Model Neurons},
	url = {https://www.sciencedirect.com/science/article/pii/S0006349572860685},
	volume = {12},
	year = {1972}}

@article{Wilson1973,
	author = {Wilson, H. R. and Cowan, J. D.},
	date = {1973/09/01},
	doi = {10.1007/BF00288786},
	id = {Wilson1973},
	isbn = {1432-0770},
	journal = {Kybernetik},
	number = {2},
	pages = {55--80},
	title = {A mathematical theory of the functional dynamics of cortical and thalamic nervous tissue},
	url = {https://doi.org/10.1007/BF00288786},
	volume = {13},
	year = {1973}}

@article{Wyller2007,author = {Wyller, John and Blomquist, Patrick and Einevoll, Gaute T.},
	date = {2007/01/01/},
	doi = {https://doi.org/10.1016/j.physd.2006.10.004},
	isbn = {0167-2789},
	journal = {Physica D: Nonlinear Phenomena},
	number = {1},
	pages = {75--93},
	title = {Turing instability and pattern formation in a two-population neuronal network model},
	url = {https://www.sciencedirect.com/science/article/pii/S0167278906004039},
	volume = {225},
	year = {2007}}

@article{Budzinski2023,
	author = {Budzinski, Roberto C. and Nguyen, Tung T. and Benigno, Gabriel B. and Đo{\`a}n, Jacqueline and Min{\'a}{\v c}, J{\'a}n and Sejnowski, Terrence J. and Muller, Lyle E.},
	date = {2023/03/02/},
	day = {02},
	doi = {10.1103/PhysRevResearch.5.013159},
	id = {10.1103/PhysRevResearch.5.013159},
	j1 = {PRRESEARCH},
	journal = {Physical Review Research},
	journal1 = {Phys. Rev. Res.},
	month = {03},
	number = {1},
	pages = {013159--},
	publisher = {American Physical Society},
	title = {Analytical prediction of specific spatiotemporal patterns in nonlinear oscillator networks with distance-dependent time delays},
	url = {https://link.aps.org/doi/10.1103/PhysRevResearch.5.013159},
	volume = {5},
	year = {2023}}

@article{Roberts2019,author = {Roberts, James A. and Gollo, Leonardo L. and Abeysuriya, Romesh G. and Roberts, Gloria and Mitchell, Philip B. and Woolrich, Mark W. and Breakspear, Michael},
	date = {2019/03/05},
	doi = {10.1038/s41467-019-08999-0},
	id = {Roberts2019},
	isbn = {2041-1723},
	journal = {Nature Communications},
	number = {1},
	pages = {1056},
	title = {Metastable brain waves},
	url = {https://doi.org/10.1038/s41467-019-08999-0},
	volume = {10},
	year = {2019}}

@article{Cakan2020,
	author = {Cakan, Caglar and Obermayer, Klaus},
	date = {2020/04/23},
	journal = {PLOS Computational Biology},
	journal1 = {PLOS Computational Biology},
	month = {04},
	n2 = {Author summary Weak electrical inputs to the brain in vivo using transcranial electrical stimulation or in isolated cortex in vitro can affect the dynamics of the underlying neural populations. However, it is poorly understood what the exact mechanisms are that modulate the activity of neural populations as a whole and why the responses are so diverse in stimulation experiments. Despite this, electrical stimulation techniques are being developed for the treatment of neurological diseases in humans. To better understand these interactions, it is often necessary to simulate and analyze very large networks of neurons, which can be computationally demanding. In this theoretical paper, we present a reduced model of coupled neural populations that represents a piece of cortical tissue. This efficient model retains the dynamical properties of the large network of neurons it is based on while being several orders of magnitude faster to simulate. Due to the biophysical properties of the neuron model, an electric field can be coupled to the population. We show that weak electric fields often used in stimulation experiments can lead to entrainment of neural oscillations on the population level, and argue that the responses critically depend on the dynamical state of the neural system.},
	number = {4},
	pages = {e1007822--},
	publisher = {Public Library of Science},
	title = {Biophysically grounded mean-field models of neural populations under electrical stimulation},
	type = {doi:10.1371/journal.pcbi.1007822},
	url = {https://doi.org/10.1371/journal.pcbi.1007822},
	volume = {16},
	year = {2020}}

@article{Mironenko2021,author = {Mironenko, Andrei and Zachariae, Ulrich and de Groot, Bert L. and Kopec, Wojciech},
	date = {2021/08/20/},
	doi = {https://doi.org/10.1016/j.jmb.2021.167002},
	isbn = {0022-2836},
	journal = {Journal of Molecular Biology},
	journal1 = {Ion channels : Intersection of Structure, Function, and Pharmacology},
	number = {17},
	pages = {167002},
	title = {The Persistent Question of Potassium Channel Permeation Mechanisms},
	url = {https://www.sciencedirect.com/science/article/pii/S0022283621002035},
	volume = {433},
	year = {2021}}

@article{Combe2021Review,author = {Combe, Crescent L. and Gasparini, Sonia},
	date = {2021/11/01/},
	doi = {https://doi.org/10.1016/j.pbiomolbio.2021.06.002},
	isbn = {0079-6107},
	journal = {Progress in Biophysics and Molecular Biology},
	journal1 = {1979-2019: a ``Funny''journey lasting 40 years, the multiple roles of f/HCN channels},
	pages = {119--132},
	title = {Ih from synapses to networks: HCN channel functions and modulation in neurons},
	url = {https://www.sciencedirect.com/science/article/pii/S0079610721000626},
	volume = {166},
	year = {2021}}

@article{MartinPedersen2024,
	author = {Martin, Matteo and Pedersen, Morten Gram},
	date = {2024/03/22},
	journal = {PLOS Computational Biology},
	journal1 = {PLOS Computational Biology},
	month = {03},
	n2 = {Author summary Neurons use the frequency of electrical signals called action potentials to encode information, and various messenger systems interact with ion channels to control this so-called firing frequency. Recent experimental recordings show that the intracellular messenger cAMP can induce mixed-mode oscillations (MMOs) consisting of small-amplitude, subthreshold oscillations separating action potentials, which lowers the firing frequency greatly. We extend a recent mathematical model of neuronal electrical activity to investigate how MMOs occur from interactions between ion channels regulated by cAMP. Our simulations reproduce a range of experimental results, including cAMP-induced MMOs. We explain the model dynamics using modern geometrical methods that exploit the different timescales in the model. Our analyses show that the very slow dynamics of cAMP-regulated HCN and M ion channels is not crucial for creating MMOs, but rather that the cAMP-induced increase in their average activity is important. Our analyses suggest that both HCN and M channels are crucial for MMOs and controlling the firing frequency, which has implications for our understanding of how astrocytes control neuronal information processing. Moreover, our study raises new mathematical questions related to how super-slow dynamical variables modify MMOs.},
	number = {3},
	pages = {e1011559--},
	publisher = {Public Library of Science},
	title = {Modelling and analysis of cAMP-induced mixed-mode oscillations in cortical neurons: Critical roles of HCN and M-type potassium channels},
	type = {doi:10.1371/journal.pcbi.1011559},
	url = {https://doi.org/10.1371/journal.pcbi.1011559},
	volume = {20},
	year = {2024}}

@article{Giocomo2008,
	author = {Giocomo, Lisa M. and Hasselmo, Michael E.},
	date = {2008/09/17},
	doi = {10.1523/JNEUROSCI.3196-08.2008},
	journal = {The Journal of Neuroscience},
	journal1 = {J. Neurosci.},
	lp = {9425},
	month = {09},
	n2 = {Chronic recordings in the medial entorhinal cortex of behaving rats have found grid cells, neurons that fire when the rat is in a hexagonal array of locations. Grid cells recorded at different dorsal--ventral anatomical positions show systematic changes in size and spacing of firing fields. To test possible mechanisms underlying these differences, we analyzed properties of the hyperpolarization-activated cation current Ih in voltage-clamp recordings from stellate cells in entorhinal slices from different dorsal--ventral locations. The time constant of h current was significantly different between dorsal and ventral neurons. The time constant of h current correlated with membrane potential oscillation frequency and the time constant of the sag potential in the same neurons. Differences in h current could underlie differences in membrane potential oscillation properties and contribute to grid cell periodicity along the dorsal--ventral axis of medial entorhinal cortex.},
	number = {38},
	pages = {9414},
	title = {Time Constants of h Current in Layer II Stellate Cells Differ along the Dorsal to Ventral Axis of Medial Entorhinal Cortex},
	url = {http://www.jneurosci.org/content/28/38/9414.abstract},
	volume = {28},
	year = {2008}}

@article{HillTononi2005,
	author = {Hill, Sean and Tononi, Giulio},
	date = {2005/03/01},
	doi = {10.1152/jn.00915.2004},
	isbn = {0022-3077},
	journal = {Journal of Neurophysiology},
	journal1 = {Journal of Neurophysiology},
	month = {2025/07/05},
	n2 = {When the brain goes from wakefulness to sleep, cortical neurons begin to undergo slow oscillations in their membrane potential that are synchronized by thalamocortical circuits and reflected in EEG slow waves. To provide a self-consistent account of the transition from wakefulness to sleep and of the generation of sleep slow waves, we have constructed a large-scale computer model that encompasses portions of two visual areas and associated thalamic and reticular thalamic nuclei. Thousands of model neurons, incorporating several intrinsic currents, are interconnected with millions of thalamocortical, corticothalamic, and both intra- and interareal corticocortical connections. In the waking mode, the model exhibits irregular spontaneous firing and selective responses to visual stimuli. In the sleep mode, neuromodulatory changes lead to slow oscillations that closely resemble those observed in vivo and in vitro. A systematic exploration of the effects of intrinsic currents and network parameters on the initiation, maintenance, and termination of slow oscillations shows the following. 1) An increase in potassium leak conductances is sufficient to trigger the transition from wakefulness to sleep. 2) The activation of persistent sodium currents is sufficient to initiate the up-state of the slow oscillation. 3) A combination of intrinsic and synaptic currents is sufficient to maintain the up-state. 4) Depolarization-activated potassium currents and synaptic depression terminate the up-state. 5) Corticocortical connections synchronize the slow oscillation. The model is the first to integrate intrinsic neuronal properties with detailed thalamocortical anatomy and reproduce neural activity patterns in both wakefulness and sleep, thereby providing a powerful tool to investigate the role of sleep in information transmission and plasticity.},
	number = {3},
	pages = {1671--1698},
	publisher = {American Physiological Society},
	title = {Modeling Sleep and Wakefulness in the Thalamocortical System},
	type = {doi: 10.1152/jn.00915.2004},
	url = {https://doi.org/10.1152/jn.00915.2004},
	volume = {93},
	year = {2005},
	year1 = {2005}}

@article{Lee2017,
  author    = {Lee, Chia-Hsiu and MacKinnon, Roderick},
  title     = {Structures of the Human HCN1 Hyperpolarization-Activated Channel},
  journal   = {Cell},
  year      = {2017},
  volume    = {168},
  number    = {1-2},
  pages     = {111--120.e11},
  doi       = {10.1016/j.cell.2016.12.023},
  pmid      = {28086084},
  pmcid     = {PMC5496774},
  url       = {https://doi.org/10.1016/j.cell.2016.12.023},
  note      = {Epub 2017 Jan 12}
}

@article{Pape1996,
	author = {Pape, H-C},
	doi = {https://doi.org/10.1146/annurev.ph.58.030196.001503},
	issn = {1545-1585},
	journal = {Annual Review of Physiology},
	number = {Volume 58, 1996},
	pages = {299-327},
	publisher = {Annual Reviews},
	title = {Queer Current and Pacemaker: The Hyperpolarization-Activated Cation Current in Neurons},
	type = {Journal Article},
	url = {https://www.annualreviews.org/content/journals/10.1146/annurev.ph.58.030196.001503},
	volume = {58},
	year = {1996}}

@article{Biel2009,
  author    = {Biel, Martin and Wahl-Schott, Christian and Michalakis, Stylianos and Zong, Xiong},
  title     = {Hyperpolarization-activated cation channels: from genes to function},
  journal   = {Physiological Reviews},
  year      = {2009},
  volume    = {89},
  number    = {3},
  pages     = {847--885},
  doi       = {10.1152/physrev.00029.2008},
  pmid      = {19584315}
}

@article{Abbas2006,author = {Abbas, S. Y. and Ying, S. -W. and Goldstein, P. A.},
	date = {2006/01/01/},
	doi = {https://doi.org/10.1016/j.neuroscience.2006.05.034},
	isbn = {0306-4522},
	journal = {Neuroscience},
	number = {4},
	pages = {1811--1825},
	title = {Compartmental distribution of hyperpolarization-activated cyclic-nucleotide-gated channel 2 and hyperpolarization-activated cyclic-nucleotide-gated channel 4 in thalamic reticular and thalamocortical relay neurons},
	url = {https://www.sciencedirect.com/science/article/pii/S0306452206007251},
	volume = {141},
	year = {2006}}

@article{McCormick1992,
	author = {McCormick, David A. },
	date = {1992/10/01/},
	doi = {https://doi.org/10.1016/0301-0082(92)90012-4},
	isbn = {0301-0082},
	journal = {Progress in Neurobiology},
	number = {4},
	pages = {337--388},
	title = {Neurotransmitter actions in the thalamus and cerebral cortex and their role in neuromodulation of thalamocortical activity},
	url = {https://www.sciencedirect.com/science/article/pii/0301008292900124},
	volume = {39},
	year = {1992}}

@article{Bhattacharjee2005,
	author = {Bhattacharjee, Arin and Kaczmarek, Leonard K.},
	date = {2005/08/01},
	doi = {10.1016/j.tins.2005.06.003},
	isbn = {0166-2236},
	journal = {Trends in Neurosciences},
	journal1 = {Trends in Neurosciences},
	month = {2025/09/16},
	number = {8},
	pages = {422--428},
	publisher = {Elsevier},
	title = {For K$^{+}$ channels, Na$^{+}$ is the new Ca$^{2+}$},
	type = {doi: 10.1016/j.tins.2005.06.003},
	url = {https://doi.org/10.1016/j.tins.2005.06.003},
	volume = {28},
	year = {2005},
	year1 = {2005}}

@article{Jones2003,
  author    = {Jones, B. E.},
  title     = {Arousal Systems},
  journal   = {Frontiers in Bioscience},
  year      = {2003},
  volume    = {8},
  pages     = {s438--s451},
  doi       = {10.2741/1074},
  pmid      = {12700104}
}

@article{DestexheBabloyantz1993,
	author = {Destexhe, A. and Babloyantz, A.},
	id = {00001756-199302000-00028},
	isbn = {0959-4965},
	journal = {NeuroReport},
	n2 = {WE investigated the kinetic properties of the hyperpolarization-activated inward current (Ih) of thalamocortical (TC) neurons. Recently, it was shown that this current is characterized by different time constants of activation and inactivation, which was in apparent conflict with the single-exponential time course of the current. We introduce here a model of Ib based on the cooperation of a slow and a fast activation variable and show that this kinetic scheme accounts for these apparently conflicting experimental data. We also report that following the combination of such a current with other currents seen in TC cells, one observes several types of oscillating behavior, similar to the slow oscillations and the spindle-like oscillations seen in vitro.  {\copyright} Lippincott-Raven Publishers.},
	number = {2},
	title = {A model of the inward current Ih and its possible role in thalamocortical oscillations},
	url = {https://journals.lww.com/neuroreport/fulltext/1993/02000/a_model_of_the_inward_current_ih_and_its_possible.28.aspx},
	volume = {4},
	year = {1993}}

@article{Destexhe1993,author = {Destexhe, A. and McCormick, D. A. and Sejnowski, T. J.},
	date = {1993/12/01/},
	doi = {https://doi.org/10.1016/S0006-3495(93)81297-9},
	isbn = {0006-3495},
	journal = {Biophysical Journal},
	number = {6},
	pages = {2473--2477},
	title = {A model for 8--10 Hz spindling in interconnected thalamic relay and reticularis neurons},
	url = {https://www.sciencedirect.com/science/article/pii/S0006349593812979},
	volume = {65},
	year = {1993}}

@article{Budzinskiy2020,
    author = {Budzinskiy, S. and Beuter, A. and Volpert, V.},
    title = {Nonlinear analysis of periodic waves in a neural field model},
    journal = {Chaos: An Interdisciplinary Journal of Nonlinear Science},
    volume = {30},
    number = {8},
    pages = {083144},
    year = {2020},
    month = {08},issn = {1054-1500},
    doi = {10.1063/5.0012010},
    url = {https://doi.org/10.1063/5.0012010},
}

@article{Erazo-Toscano2023,
	author = {Erazo-Toscano, Ricardo and Osan, Remus},
	date = {2023/03/15/},
	day = {15},
	doi = {10.1103/PhysRevE.107.034403},
	id = {10.1103/PhysRevE.107.034403},
	j1 = {PRE},
	journal = {Physical Review E},
	journal1 = {Phys. Rev. E},
	month = {03},
	number = {3},
	pages = {034403--},
	publisher = {American Physical Society},
	title = {Synaptic propagation in neuronal networks with finite-support space-dependent coupling},
	url = {https://link.aps.org/doi/10.1103/PhysRevE.107.034403},
	volume = {107},
	year = {2023}}

@article{Omelchenko2024,
	author = {Omel'chenko, Oleh E. and Laing, Carlo R.},
	date = {2024/09/25/},
	day = {25},
	doi = {10.1103/PhysRevE.110.034411},
	id = {10.1103/PhysRevE.110.034411},
	j1 = {PRE},
	journal = {Physical Review E},
	journal1 = {Phys. Rev. E},
	month = {09},
	number = {3},
	pages = {034411--},
	publisher = {American Physical Society},
	title = {Activity patterns in ring networks of quadratic integrate-and-fire neurons with synaptic and gap junction coupling},
	url = {https://link.aps.org/doi/10.1103/PhysRevE.110.034411},
	volume = {110},
	year = {2024}}

@article{Quabbaj2009,author = {Qubbaj, Murad R. and Jirsa, Viktor K.},
	date = {2009/12/01/},
	doi = {https://doi.org/10.1016/j.physd.2009.09.014},
	isbn = {0167-2789},
	journal = {Physica D: Nonlinear Phenomena},
	number = {23},
	pages = {2331--2346},
	title = {Neural field dynamics under variation of local and global connectivity and finite transmission speed},
	url = {https://www.sciencedirect.com/science/article/pii/S0167278909002887},
	volume = {238},
	year = {2009}}

@article{Bressloff2010,
	author = {Bressloff, Paul C. },
	date = {2010/11/03/},
	day = {03},
	doi = {10.1103/PhysRevE.82.051903},
	id = {10.1103/PhysRevE.82.051903},
	j1 = {PRE},
	journal = {Physical Review E},
	journal1 = {Phys. Rev. E},
	month = {11},
	number = {5},
	pages = {051903--},
	publisher = {American Physical Society},
	title = {Metastable states and quasicycles in a stochastic {W}ilson-{C}owan model of neuronal population dynamics},
	url = {https://link.aps.org/doi/10.1103/PhysRevE.82.051903},
	volume = {82},
	year = {2010}}

@article{Ermentrout1998,author = {Ermentrout, Bard},
	date = {1998/05/01},
	doi = {10.1023/A:1008822117809},
	id = {Ermentrout1998},
	isbn = {1573-6873},
	journal = {Journal of Computational Neuroscience},
	number = {2},
	pages = {191--208},
	title = {The Analysis of Synaptically Generated Traveling Waves},
	url = {https://doi.org/10.1023/A:1008822117809},
	volume = {5},
	year = {1998}}

@article{Kilpatrick2010,author = {Kilpatrick, Zachary P. and Bressloff, Paul C.},
	date = {2010/05/01/},
	doi = {https://doi.org/10.1016/j.physd.2009.06.003},
	isbn = {0167-2789},
	journal = {Physica D: Nonlinear Phenomena},
	journal1 = {Mathematical Neuroscience},
	number = {9},
	pages = {547--560},
	title = {Effects of synaptic depression and adaptation on spatiotemporal dynamics of an excitatory neuronal network},
	url = {https://www.sciencedirect.com/science/article/pii/S0167278909001833},
	volume = {239},
	year = {2010}}

@article{Amari1977,
  author  = {Amari, S.},
  title   = {Dynamics of pattern formation in lateral-inhibition type neural fields},
  journal = {Biological Cybernetics},
  year    = {1977},
  volume  = {27},
  number  = {2},
  pages   = {77--87},
  doi     = {10.1007/BF00337259},
}

@article{Folias2004,
    author = {Folias, Stefanos E. and Bressloff, Paul C.},
	doi = {10.1137/030602629},
	journal = {SIAM Journal on Applied Dynamical Systems},
	number = {3},
	pages = {378-407},
	title = {Breathing Pulses in an Excitatory Neural Network},
	url = {https://doi.org/10.1137/030602629},
	volume = {3},
	year = {2004}}

@article{Stroemsdoerfer-preprint,
        title={Spike-frequency and h-current based adaptation are dynamically equivalent in a {W}ilson-{C}owan field model}, 
        author={Ronja Strömsdörfer and Klaus Obermayer},
        year={2025},
        journal={arXiv},
        eprint={2510.08436},
        archivePrefix={arXiv},
        primaryClass={nlin.AO},
        url={https://arxiv.org/abs/2510.08436}, 
}

@article{Marino2025,
    author = {Marino, Raffaele and Buffoni, Lorenzo and Chicchi, Lorenzo and Patti, Francesca Di and Febbe, Diego and Giambagli, Lorenzo and Fanelli, Duccio},
    title = {Learning in Wilson-Cowan Model for Metapopulation},
    journal = {Neural Computation},
    volume = {37},
    number = {4},
    pages = {701-741},
    year = {2025},
    month = {03},
    issn = {0899-7667},
    doi = {10.1162/neco_a_01744},
}

@article{Trabocchi2025,
      title={Generalized Wilson-Cowan model with short term synaptic plasticity}, 
      author={Tommaso Trabocchi and Raffaella Burioni and Lucilla de Arcangelis and Duccio Fanelli},
      year={2025},
      journal={arXiv},
      eprint={2511.16252},
      archivePrefix={arXiv},
      primaryClass={cond-mat.dis-nn},
      url={https://arxiv.org/abs/2511.16252},
}


\end{document}